\documentclass[12pt]{article}
\usepackage{amssymb,amsmath}
\makeatletter

\@addtoreset{equation}{section}
\def\section{\@startsection {section}{1}{\z@}{-2.5ex plus -1ex minus
 -.2ex}{1.3ex plus .2ex}{\large\bf}}
\def\subsection{\@startsection{subsection}{2}{\z@}{-2.25ex plus%
 -1ex minus -.2ex}{0.5ex plus .2ex}{\bf}}

\def\Ad{\mathrm{Ad}}

\def\ba{{\mbox{\boldmath $a$}}}
\def\bx{{\mbox{\boldmath $x$}}}
\def\bs{{\mbox{\boldmath $s$}}}

\def\bj{{\mbox{\boldmath $j$}}}
\def\bk{{\mbox{\boldmath $k$}}}

\def\bq{{\mbox{\boldmath $q$}}}

\newcommand{\gothf}{\mathcal F }
\newcommand{\gothg}{\mathfrak g }
\newcommand{\gothn}{\mathfrak n }
\newcommand{\gothh}{\mathfrak h }
\newcommand{\gothe}{\mathfrak e }

\newcommand{\RR}{\mathbb{R}}
\newcommand{\CC}{\mathbb{C}}

\def\bea{\begin{eqnarray}}
\def\eea{\end{eqnarray}}
\newtheorem{theorem}{Theorem}[section]

\newtheorem{lemma}[theorem]{Lemma}

\newtheorem{definition}[theorem]{Definition}

\def\bmz{\left(\begin{array}{2,2}}
\def\emz{\end{array}\right)}
\def\bmd{\left(\begin{array}{3,3}}
\def\emd{\end{array}\right)}

\newcommand{\ad}[0] {\mathrm{ad}}

\newcommand{\inv}[0]{{-1}}

\newcommand{\mi}[0]{{M_i}}
\newcommand{\ai}[0]{{A_i}}
\newcommand{\bi}[0]{{B_i}}

\newcommand{\mip}[0]{{M'_i}}
\newcommand{\aip}[0]{{A'_i}}
\newcommand{\bip}[0]{{B'_i}}
\newcommand{\mjp}[0]{{M'_j}}
\newcommand{\ajp}[0]{{A'_j}}
\newcommand{\bjp}[0]{{B'_j}}

\newcommand{\me}[0]{{M_1}}
\newcommand{\aee}[0]{{A_1}}
\newcommand{\bee}[0]{{B_1}}

\newcommand{\mf}[0]{{M_n}}
\newcommand{\af}[0]{{A_g}}
\newcommand{\bff}[0]{{B_g}}

\newcommand{\oo}[0]{\otimes}

\newcommand{\cif}[0]{\mathcal{C}^\infty}

\newcommand{\prgr}{G\ltimes \mathfrak{g}^*}
\newcommand{\defgr}[0]{G\ltimes \mathcal{C}^\infty(G)}
\newcommand{\repind}[0]{{\mu_1s_1\ldots\mu_ns_n}}

\newcommand{\cala}[0]{\mathcal{A}}

\newcommand{\calc}[0]{\mathcal{C}}
\newcommand{\calm}[0]{\mathcal{M}}
\newcommand{\calq}[0]{\mathcal{Q}}
\newcommand{\calqq}[0]{\widehat{\mathcal{Q}}}

\newcommand{\flh}[3] {f_{{#1}{#2}}^{\;\;\;\;{#3}}\;}

\newcommand{\tenltimes}[0] {\mbox{$\subset\!\!\!\!\!\!\times$}}

\begin{document}
\parskip 6pt
\parindent 0pt
\begin{flushright}
HWM-03-20\\
EMPG-03-19\\
hep-th/0310218
\end{flushright}

\begin{center}

\baselineskip 24 pt
{\Large \bf  The quantisation of Poisson structures arising in } \\
 {\Large \bf  Chern-Simons theory with gauge group $\prgr$}

\vspace{1cm}
{\large C.~Meusburger\footnote{\tt  cm@ma.hw.ac.uk}
and    B.~J.~Schroers\footnote{\tt bernd@ma.hw.ac.uk} \\
Department of Mathematics, Heriot-Watt University \\
Edinburgh EH14 4AS, United Kingdom } \\

\vspace{0.5cm}

{ 22 October 2003}

\end{center}

\begin{abstract}

\noindent  We quantise  a Poisson structure on $H^{n+2g}$, where
$H$ is a semidirect product group of the form
$G\ltimes\mathfrak{g}^*$.
This Poisson structure
arises in the combinatorial description of the phase space
of Chern-Simons theory with
 gauge group $G\ltimes\mathfrak{g}^*$
on $\mathbb{R}\times S_{g,n}$, where $S_{g,n}$ is a surface of  genus
$g$ with $n$ punctures. The quantisation of this Poisson structure is
a key step in the quantisation of Chern-Simons theory with gauge group
 $G\ltimes\mathfrak{g}^*$.
We construct the quantum algebra and its irreducible
representations and show that  the quantum double $D(G)$ of the group
$G$ arises naturally as a symmetry of the quantum algebra.
\end{abstract}

\section{Introduction}

 The aim of this paper is the quantisation of a  Poisson
structure  which arises  in the study of Chern-Simons gauge theory with
semidirect product gauge group  $H=\prgr$, where $G$ is a
 Lie group, $\gothg^*$ the
dual of its Lie algebra $\gothg$  viewed as an abelian group
 and $G$ acts on $\gothg^*$ in the co-adjoint representation.
Such gauge groups occur in the
Chern-Simons formulation of (2+1)-dimensional gravity \cite{Witten1},
 where the gauge
group is the
three-dimensional Poincar\'e group or Euclidean group,
depending on the signature of space-time. Besides their
mathematical interest, Chern-Simons gauge theories with gauge groups
of this type are therefore of physical relevance.

The Poisson algebra studied in this article, in the following
referred to as flower algebra, was first defined by Alekseev,
Grosse and Schomerus \cite{AGSI, AGSII, AS}
in the context of an earlier work by Fock
and Rosly \cite{FR}. Fock and Rosly showed that it is possible to
describe the Poisson structure of the phase space of Chern-Simons
theory, the moduli
space of flat connections,
on an oriented, punctured  surface in terms of a graph embedded into this
surface. By assigning a classical $r$-matrix for the gauge group  to
each vertex of the graph, they define a Poisson structure on the space
of graph connections. After Poisson reduction with respect to
graph gauge transformations, this Poisson structure agrees with
the canonical Poisson
structure on the moduli space \cite{Goldman,AB}. Alekseev, Grosse and Schomerus
specialised this description to a set of curves representing the
generators of the surface's fundamental group as a
particularly simple graph. They then obtain a Poisson structure on the
space of holonomies associated to these generating curves, which -
because  this set of curves resembles  a flower -
we will call  the flower algebra.
Via Poisson  reduction
with respect to simultaneous conjugation of the holonomies
with elements of the gauge group, it induces
the canonical Poisson structure on the moduli space.
Although the case of  surfaces with a boundary is more
involved due to additional degrees of freedom arising at the boundary,
 the flower algebra still remains an important ingredient.

The relevance of the flower algebra for the phase space of
Chern-Simons
gauge theory makes its quantisation an important
task. For the case of compact, semisimple Lie groups this has been
achieved by Alekseev, Grosse and
Schomerus \cite{AGSI, AGSII, AS} with their formalism of
combinatorial quantisation of
Chern-Simons gauge theories. However, the case of (non-compact and
  non-semisimple) Lie groups of type $H=\prgr$ such as the three-dimensional
Poincar\'e group arising in (2+1)-dimensional gravity
is  less well investigated. In addition to the need to establish a quantisation procedure, the physical relevance of  this case also calls for a more explicit description in terms of coordinates with a direct physical meaning.

In this article we show that this can be achieved for groups of type
$H=\prgr$,
 where $G$ is a finite-dimensional, connected,
simply connected and  unimodular
Lie group. The assumptions of simply-connectedness and unimodularity
are made for convenience;  dropping them would lead to
technical modifications without affecting the essence of our results.
By extending and adapting
the work of Alekseev and Malkin \cite{AMII} to this case, we construct a
bijective decoupling transformation which breaks up the Poisson
structure of the flower algebra  into a set of Poisson-commuting
building blocks, a copy of the dual Poisson-Lie group $H^*$ for
each puncture and a copy of its Heisenberg double $D_+(H)$ for each handle. We
 quantise these building blocks and then define a quantum counterpart of the decoupling
transformation to construct the quantum algebra for the original
Poisson structure and its irreducible Hilbert space
representations. After investigating the action of the
quantum symmetries on the
representation spaces, we relate them to the quantum double $D(G)$ of the group $G$.

The article is structured as follows. In Sect.~2 we establish the
relevant definitions and notations and discuss various Poisson
structures associated to  groups $\prgr$ that are relevant for our
 description of the flower algebra. Extending our treatment
 \cite{we}  of the
 universal cover $\widetilde{SO(2,1)}\ltimes\mathbb{R}^3$ of the
 Poincar\'e group in three dimensions, we introduce the
 flower algebra on a genus $g$  surface with $n$ punctures as
 defined in \cite{AGSI, AGSII, AS} and give an explicit description of its Poisson
 structure for groups of type $H=\prgr$. We define a bijective
 decoupling transformation that maps this Poisson
 structure onto the direct sum of $n$ copies of the
 dual Poisson-Lie group $H^*$ and $g$ copies of the Heisenberg double
 $D_+(H)$. Finally, we show that elements of the
 semidirect product group $G\ltimes\cif(G)$
  act as
Poisson isomorphisms on the flower
 algebra and
relate this group action to the action of the group $H$ by global conjugation.

Sect.~3 describes the quantisation of the flower algebra. Starting
from the decoupled Poisson structure, we construct the quantum algebra
and its irreducible
representations for each of the building blocks. We then define a quantum counterpart of
the classical decoupling transformation to obtain a quantisation of the original brackets of the
flower algebra.

In Sect.~4  we discuss symmetries acting on the quantised flower
algebra. We determine how  the group $G\ltimes\cif(G)$ acts on the
quantum algebra and how this action can be implemented as an
action on the representation spaces. We establish the relation between
this quantum symmetry and the quantum double
$D(G)$ of the Lie group $G$.

Sect.~5 contains our outlook and conclusions.

\section{The classical Poisson structure}

\subsection {Poisson structures associated to $\prgr$ as a Poisson Lie group}

We consider groups $H=\prgr$ which are the
semidirect product of a
 Lie group $G$ and the dual $\gothg^*$ of its  Lie algebra
$\gothg=\text{Lie}\,G$, viewed as an abelian group.
The group $G$ is assumed to be connected, simply connected
and unimodular.
Following the conventions of \cite{MaRa},
 we define $\Ad^*(g)$ to be the algebraic dual of $\Ad(g)$, i.~e.~
\bea
\label{coadact}
\langle \Ad^*(g)\bj, \xi\rangle=\langle \bj, \Ad(g)\xi \rangle\qquad\forall \bj\in\gothg^*, \xi\in\gothg, g\in G,
\eea
so that the coadjoint action of $g\in G$ is given by $\Ad^*(g^\inv)$.
Writing elements $h\in H$ as $(u,\ba)$,
 the group multiplication is given by
\bea
\label{groupmult}
(u_1,\ba_1)\cdot(u_2,\ba_2)=(u_1\cdot u_2,\ba_1+\Ad^*(u_1^\inv)\ba_2).
\eea
Often we use the parametrisation
\bea
\label{gparam}
(u,\ba)=(u,-\Ad^*(u^\inv)\bj)\qquad\text{with}\; u\in G,\;\ba,\bj\in\gothg^*,
\eea
which will be interpreted geometrically further below.

 All Lie algebras are
considered  over $\mathbb{R}$ unless stated otherwise,
and Einstein summation convention is used throughout the paper.
Let  $J_a$, $P^a$, $a=1,\ldots,\text{dim}\,G$, denote the
 generators of the Lie algebra $\gothh=\text{Lie}\,
H=\gothg\oplus\gothg^*$, such that the generators $J_a$,
$a=1,\ldots,\text{dim}\,G$, generate $\gothg=\text{Lie}\,G$ and
$P^a$, $a=1,\ldots,\text{dim}\,G$, generate $\gothg^*$. The
commutator is then given by
\bea
\label{commutator}
[J_a,J_b]=\flh a b c J_c\qquad[J_a,P^b]=-\flh a c b P^ c\qquad[P^a,P^b]=0,
\eea
where $\flh a b c$ are the structure constants of $\gothg$. The Lie algebra $\gothh$ admits the non-degenerate bilinear form
\bea
\label{blf}
\langle J_a,\, J_b\rangle=0\qquad\langle P^a,P^b\rangle=0
\qquad\langle J_a,\,P^b\rangle=\delta_a^b,
\eea
which is $H$-invariant by virtue of \eqref{coadact}.
We may view $\gothh$ as the classical double of the
Lie bialgebra $\gothg$ with standard commutator and trivial
cocommutator, where the pairing between $\gothg$ and $\gothg^*$
is given by \eqref{blf}. It has a coboundary
Lie bialgebra structure with commutator \eqref{commutator} and
cocommutator $\delta:\gothh\rightarrow\gothh\otimes\gothh$
\bea
\label{cocomm}
\delta(J_a)=0\qquad\delta(P^a)=\flh b c a P^b\otimes P^c,
\eea
which arises from the classical  $r$-matrix
\begin{align}
\label{rmatrix}
r=P^a\otimes J_a\in\gothh\otimes\gothh.
\end{align}
This Lie bialgebra structure on the space
$\gothh=\gothg\oplus\gothg^*$ is the infinitesimal version, the
tangent Lie bialgebra, of an associated Poisson-Lie structure on the
group $H$.
If we denote by
 $\tilde P_L^{a},\tilde P_R^{a},\tilde
J^L_a,\tilde J^R_a$,
$a=1,\ldots,\text{dim}\,G$,  the right- and left-invariant
vector fields on $H$
\begin{align}
\label{vecfields}
&\tilde P_R^{a} f(u,-\Ad^*(u^\inv)\bj)=\frac{\text{d}}{\text{d}t}|_{t=0}f\left((u,-\Ad^*(u^\inv)\bj)\cdot tP^a\right)=-\frac{\partial f}{\partial j_a}(u,-\Ad^*(u^\inv)\bj)\nonumber\\
&\tilde P^a_L f(u,-\Ad^*(u^\inv)\bj)=\frac{\text{d}}{\text{d}t}|_{t=0}f\left(-tP^a\cdot (u,-\Ad^*(u^\inv)\bj)\right)=\Ad^*(u)_b^{\;\;a}\frac{\partial f}{\partial j_b}(u,-\Ad^*(u^\inv)\bj)\nonumber\\
&\tilde J_a^R f(u,-\Ad^*(u^\inv)\bj)=\frac{\text{d}}{\text{d}t}|_{t=0}f\left( (u,-\Ad^*(u^\inv)\bj)\cdot e^{tJ_a}\right)\nonumber\\
&\qquad\qquad\qquad\qquad\;\;\;\,=J_a^R f(u,-\Ad^*(u^\inv)\bj)+ \flh a b c \frac{\partial f}{\partial j_b}(u,-\Ad^*(u^\inv)\bj)j_c\nonumber\\
&\tilde J^L_a f(u,-\Ad^*(u^\inv)\bj)= J_a^L f( u,-\Ad^*(u^\inv)\bj),
\end{align}
with $\bj=j_a P^a$, $f\in\calc^\infty(H)$ and
the left-and right-invariant vector fields $J_a^R$, $J_a^L$ on $G$
\begin{align}
\label{gvecfields}
&J_a^Rf:=\frac{\text{d}}{\text{d}t}|_{t=0}f(u e^{tJ_a})\qquad
J_a^Lf:=\frac{\text{d}}{\text{d}t}|_{t=0}f(e^{-tJ_a}u) & &\text{for}\, f\in\cif(G),
\end{align} this Poisson-Lie structure is given by the Poisson bivector
\bea
\label{grbivect}
B_H=\tilde P^a_L\wedge \tilde J^L_a-\tilde P^a_R\wedge \tilde J_a^{R}.
\eea
Similarly, there is a Poisson-Lie group structure associated to the dual
$\gothh^*$ of the Lie bialgebra $\gothh=\gothg\oplus\gothg^*$,
the dual $H^*$ of the Poisson-Lie group $H$. As a group, it is the direct product
$G\times\gothg^*$ with group multiplication
$(u_1,\bj_1)\cdot(u_2,\bj_2)=(u_1u_2,\bj_1+\bj_2)$. The
global diffeomorphism $H^*\rightarrow H$, $(u,\bj)\mapsto
(u,-\Ad^*(u^\inv)\bj)$ \cite{CP} allows us to describe its
Poisson structure in terms
of the following Poisson bivector on $H$
\bea
\label{conjbivect}
B_{H^*}=\frac{1}{2}\left(\tilde P^a_L\wedge \tilde J_a^{L}+ \tilde
  P^a_R\wedge \tilde J_a^{R}\right)+\tilde P^a_R\wedge \tilde J_a^{L}.
\eea
A more explicit formula in terms of the parametrisation \eqref{gparam} is given in Sect.~\ref{decoupsect}, \eqref{partbr}.
The
symplectic leaves of this Poisson structure are the conjugacy classes
in $H$ \cite{AMI, setishan}. As they play an important role in the
quantisation of the flower algebra, we need to introduce some
additional notation that will allow us to take a more geometric
viewpoint and relate them to the conjugacy classes in the group $G$.
For an element $u\in G$ let  $N_u=\{n\in G|nu
n^{-1}=u\}$ denote its stabiliser group with Lie algebra  $\gothn_u$ and dimension
 $\nu_u$. Pick a basis $\{J_\alpha\}$, $\alpha=1,\ldots,\nu_u$ of $\gothn_u$
and complete it to a basis $\{J_a\}$,
$a=1,\ldots,\text{dim}\,G$, of $\gothg$. If we denote the dual basis of
$\gothg^*$ by $\{P_a\}$, $a=1,\ldots,\text{dim}\,G$, as above and
define $\gothn^*_u =\mbox{Span}(P_1,\ldots,P_{\nu_u})$, we
have the decomposition
\bea
\label{decomp}
\gothg^* = \mbox{Im}(1- \Ad^*(u))\oplus \gothn_u^*\quad\text{for
  each}\; u\in G.
\eea
This decomposition gives rise to a convenient parametrisation of the
conjugacy classes in the group  $H$ that clarifies their relation to
the conjugacy classes in $G$. From a fixed element $(g,-\Ad^*(g^{-1})\bs)\in H$
  other elements $(u, -\Ad^*(u^{-1})\bj)$ in the same conjugacy class
  in $H$
are obtained by conjugation with $(v,\bx)\in H$ and explicitly given
by
\bea
\label{conjparam}
u &=& vgv^{-1} \\
\bj&=&\Ad^*(v^{-1})\bs + (1-\Ad^*(u))\bx. \nonumber
\eea
Equations \eqref{decomp}, \eqref{conjparam} allow us to characterise a
 conjugacy class in $H$ by
 picking a group element $g_\mu\in G$ and an element $\bs \in
 \gothn_{g_\mu}^*$ in the dual Lie algebra of its stabiliser. After
 choosing $g_\mu\in G$, the remaining arbitrariness in the choice of
 $\bs$ is parametrised by elements  $n\in N_{g_\mu}$. Under their
action $\bs $ sweeps out a co-adjoint orbit ${\cal O}_s$ in
$\gothn^*_{g_\mu}$. Conjugacy classes $\calc_{\mu s}$ in $H$
are therefore uniquely characterised by $G$-conjugacy classes $\calc_\mu=\{v g_\mu v^\inv\,|\, v\in G\}$
 and co-adjoint orbits ${\cal O}_s$ in
$\gothn^*_{g_\mu}$.  With respect to fixed $g_\mu\in G$ and $\bs \in
 \gothn_{g_\mu}^*$, they are given as the image of the map
\bea
\label{conjmap}
\mbox{conj}_{\mu s}: H &\rightarrow& \calc_{\mu s} \\
(v,\bx)& \mapsto & (u,\bj) =
(v,\bx)(g_{\mu},-\bs)(v,\bx)^{-1}\nonumber.
\eea
In geometric terms, this amounts to the following. The identification
$T_u^* \calc_\mu = \mbox{Im}(1-\Ad^*(u))$ allows us to write the
$H$-conjugacy classes $C_{\mu s}$ locally as the product
of the cotangent bundle  $T^* \calc_\mu$ and ${\cal O}_s$.
With the projection $\pi_u:\gothg^* \mapsto \gothg^*/\gothn_u^*\simeq
\mbox{Im}(1-\Ad^*(u))$
we then have the bundle
\bea
\calc_{\mu s} &\rightarrow & T^* \calc_\mu \\
(u,-\Ad^*(u^{-1})\bj)&\mapsto& (u,(-\Ad^*(u^{-1}))\pi_u(\bj))\nonumber
\eea
with typical fibre ${\cal O}_s$.

The third Poisson structure associated to the Lie bialgebra $\gothh$
that we will be relevant in this article is the so-called Heisenberg double $D_+(H)$
of the Poisson-Lie group $H$. It is the Poisson structure on the direct
product $H\times H$ defined by the following Poisson bivector
\begin{align}
\label{heisdouble}
B_{D_+(H)}=&\frac{1}{2}\left(\tilde P^a_{R_1}\wedge \tilde
  J_a^{R_1}+\tilde P^a_{L_1}\wedge \tilde J_a^{L_1}+\tilde
  P^a_{R_2}\wedge \tilde J_a^{R_2}+\tilde P^a_{L_2}\wedge \tilde J_a^{L_2}\right)\\
&+\tilde P^a_{R_1}\wedge \tilde  J_a^{R_2}+\tilde P^a_{L_1}\wedge
  \tilde J_a^{L_2}\nonumber,
\end{align} where $\tilde P^a_{R_i}$, $\tilde P^a_{L_i}$, $\tilde J^{R_i}_a$,
  $\tilde J^{L_i}_a$ denote the left-and right-invariant vector fields on the
  two copies of $H$. In \cite{setishan, wu} it was shown
that this Poisson structure
is symplectic; however, it is {\em not} a Poisson-Lie structure.
We derive an explicit formula for this Poisson structure in Sect.~\ref{decoupsect},\eqref{hbr} and show that
  in suitable coordinates it is the canonical Poisson structure of
  the cotangent bundle $T^*(G\times G)$.

\subsection{The flower algebra}
\label{flowalg}
After introducing the relevant concepts and definitions,
we are now ready to discuss the flower algebra for semidirect product
gauge groups $\prgr$ on a genus $g$  surface $S_{g,n}$
with $n$ punctures.

The phase space of Chern-Simons theory on the surface $S_{g,n}$
(with any gauge group) is the
moduli space of flat connections. It
can be described in terms of a graph embedded into the
surface \cite{FR}. The moduli space as a manifold with singularities
is then obtained as
the quotient of the space of flat graph connections modulo graph gauge
transformations. However, as the canonical Poisson structure of the
underlying Chern-Simons gauge theory does in general {\em not} induce
a Poisson structure on the space of graph connections, this
description can a priori not provide a description of the canonical
Poisson structure on the moduli space. As explained in the
introduction,
 Fock and Rosly \cite{FR}
succeeded in defining a (non-canonical)  Poisson structure on the space
of graph connections that induces the canonical Poisson structure on
the moduli space.
Alekseev, Grosse and Schomerus
\cite{AGSI, AGSII, AS} specialised this description to the simplest
graph
that can be used to describe the underlying  surface: a set of
curves representing the generators of its fundamental group.
The space of graph connections is then simply the set of
holonomies along these curves, and graph gauge transformations
act on the holonomies via simultaneous conjugation.
The resulting Poisson structure on the space of holonomies is
the flower algebra.

The case of  surfaces with boundary is more involved, as
gauge transformations that are nontrivial at the boundary acquire a
physical meaning and are no longer divided out of the phase
space. Depending on the boundary conditions imposed, there are
additional degrees of freedom associated to the boundary which enter
into the phase space. The Poisson structure then contains a
contribution of these boundary degrees of freedom as well as a bulk
term representing the internal degrees of freedom, subject to
constraints relating the two contributions.

We can now define the flower algebra, summarising the
results and definitions of \cite{AGSI, AGSII, AS}.
The first ingredient is a set of
generators for the fundamental group of the underlying Riemann surface.
The fundamental group $\pi_1(S_{g,n})$ of a genus $g$  surface
$S_{g,n}$ with $n$ punctures is generated by the equivalence classes
of a loop $m_i$, $i=1,\ldots,n$, around each puncture and two curves
$a_j$, $b_j$, $j=1,\ldots,g$ for each handle, see Fig.~1.

\vbox{
\vskip .1in
\input epsf
\epsfxsize=9 truecm
\centerline{
\epsfbox{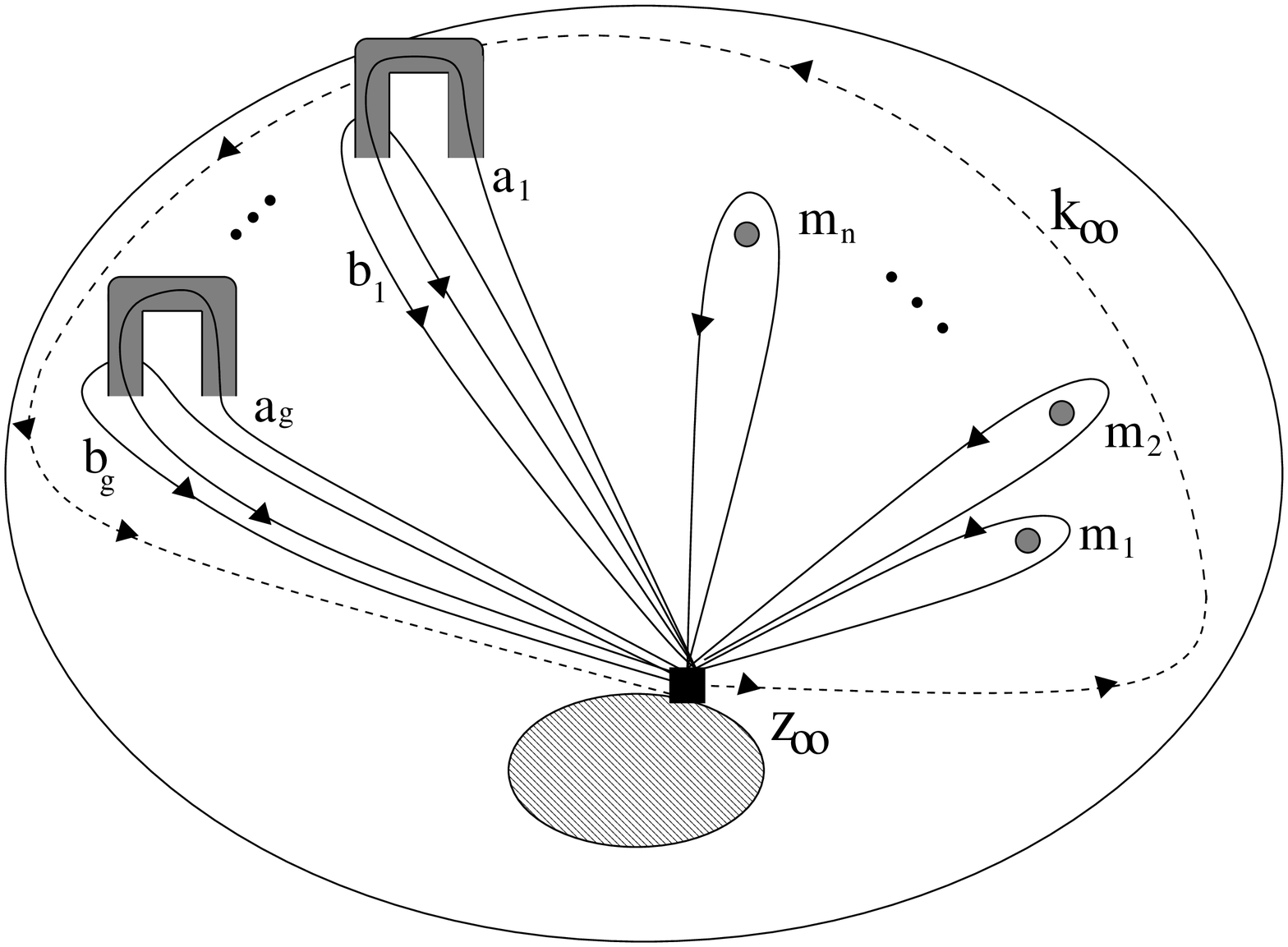}
}
\bigskip
{
\centerline{\bf Fig.~1 }
}
\centerline{\footnotesize The generators of the
fundamental group of the surface $S_{g,n}\!\setminus\! D$ with a disc
removed (shaded)}
       }

 For the  surface $S_{g,n}$,
these generators are subject to a single, defining relation
\bea
\label{pirel}
k_\infty=[b_g,a_g^{-1}]\cdot\ldots\cdot [b_1,a_1^{-1}]
\cdot m_n\cdot\ldots\cdot m_1=1,
\eea
with group commutator $[b_i,a_i^\inv]=b_i a_i^\inv b_i^\inv a_i$. If
we remove a disc $D$ from the surface, thus obtaining the surface
$S_{g,n}\setminus D$ with connected boundary  shown in Fig.~1,
the fundamental group $\pi_1(S_{g,n}\setminus D)$ is freely generated
by the $n+2g$ curves
$m_i$, $i=1,\ldots,n$ and $a_j$, $b_j$, $j=1,\ldots,g$.
 Whereas the holonomies of the curves for
each handle are general elements of the gauge group $H$, the
holonomies corresponding to the punctures are restricted to fixed
$H$-conjugacy classes $\calc_{\mu_is_i}\subset H$.
Therefore, the
space $\cala_{g,n}$ of graph connections, or holonomies, is given by
\begin{align}
\label{grconn}
\cala_{g,n}&=\big\{(M_1,\ldots,M_n,A_1,B_1,\ldots,A_g,B_g)\in\calc_{\mu_1s_1}
\times\ldots\times\calc_{\mu_n s_n}\times H^{2g}\;| \\
&\qquad\qquad\qquad\qquad\qquad [B_g,A_g^{-1}]\cdot\ldots\cdot
[B_1,A_1^{-1}]\cdot M_n\cdot\ldots\cdot M_1=1\;\big\}\nonumber.
\end{align}
The moduli space $\calm_{g,n}$ of flat $H$-connections on a closed
surface $S_{g,n}$ is obtained from this space by dividing out
simultaneous conjugation of all holonomies by the group $H$
\bea
\label{modspace}
\calm_{g,n}=\cala_{g,n}/ \sim\;\;,
\eea
where $\sim$ denotes simultaneous conjugation by an element of the
group $H$. Following the work of Alekseev, Grosse and Schomerus
\cite{AGSI, AGSII, AS}, we then have the following definition of the
flower algebra (see
Theorem 2  in \cite{AS}).

\pagebreak

\begin{definition} (Flower algebra)
\label{flower0}

The flower algebra for gauge group $H$  on a
genus $g$  surface $S_{g,n}$ with $n$ punctures is the Poisson
algebra
$\cif(H^{n+2g})$ defined by the following Poisson bivector
\begin{align}
\label{fundbivect}
B_{FR}=&\sum_{i=1}^n r^{\alpha\beta}\left(\frac{1}{2}R_\alpha^{M_i}
\wedge R_\beta^{M_i}+\frac{1}{2}L_\alpha^{M_i}\wedge L_\beta^{M_i}+
R_\alpha^{M_i}\wedge L_\beta^{M_i}\right)\\
+&\sum_{i=1}^gr^{\alpha\beta}\bigg(\frac{1}{2}\left( R_\alpha^{A_i}
\wedge R_\beta^{A_i}+L_\alpha^{A_i}\wedge L_\beta^{A_i}+R_\alpha^{B_i}
\wedge R_\beta^{B_i}+L_\alpha^{B_i}\wedge L_\beta^{B_i}\right)\nonumber\\
&\qquad\qquad+R_\alpha^{A_i}\wedge (R_\beta^{B_i}+L_\beta^{A_i}+
L_\beta^{B_i})+R_\alpha^{B_i}\wedge (L_\beta^{A_i}+L_\beta^{B_i})+
L_\alpha^{A_i}\wedge L_\beta^{B_i}\bigg)\nonumber\\
+&\sum_{1\leq i<j\leq n}r^{\alpha\beta} (R_\alpha^{M_i}+
L_\alpha^{M_i})\wedge(R_\beta^{M_j}+L_\beta^{M_j})\nonumber\\
+&\sum_{1\leq i<j\leq g}r^{\alpha\beta}(R_\alpha^{A_i}+
L_\alpha^{A_i}+R_\alpha^{B_i}+L_\alpha^{B_i})
\wedge(R_\beta^{A_j}+L_\beta^{A_j}+R_\beta^{B_j}+L_\beta^{B_j})\nonumber\\
+&\sum_{i=1}^n\sum_{j=1}^g
r^{\alpha\beta}(R_\alpha^{M_i}+L_\alpha^{M_i})
\wedge(R_\beta^{A_j}+L_\beta^{A_j}+R_\beta^{B_j}+L_\beta^{B_j})\nonumber\,,
\end{align}
where elements of $H^{n+2g}$ are denoted by
$(M_1,\ldots,M_n,A_1,B_1,\ldots,A_g,B_g)$. The coefficients
$r^{\alpha\beta}$ are the components of a classical
$r$-matrix $r\in\gothh\otimes\gothh$ for  $H$ with respect
to a given basis $Z_\alpha$, $\alpha=1,\ldots,\text{dim}\,H$
of $\gothh$ and $L^\alpha_X$, $R^\alpha_X$, $X=M_1,\ldots,M_n,A_1,B_1,
\ldots,A_g,B_g$ the right-and left invariant vector fields
corresponding to this basis.
\end{definition}

As our notation suggests,  the elements of $H^{n+2g}$ should be
thought of as holonomies around the curves depicted in Fig.~1.
Note, however, that we neither impose   the condition \ref{pirel} nor restrict
the holonomies around the punctures in the definition of the flower
algebra.

With an expression for the classical $r$-matrix and the
right-and left invariant vector fields on the different copies of $H$,
formula \eqref{fundbivect} determines the Poisson brackets of two
functions in  $\cif(H^{n+2g})$. However, in the case of groups of
type $H=\prgr$, there is an advantage in working with a slightly
different definition of the flower algebra. We expand the
vector $\bj$ in \eqref{gparam} as $\bj=j_bP^b$, and denote by the same
symbol the maps $j_a\in\cif(H): (u,-\Ad^*(u^\inv)\bj)\mapsto
j_a$. Instead of  $\cif(H)$ we then consider the algebra
generated  by the functions
in $\cif(G)$ together with these maps $j_a$. By inserting
the $r$-matrix \eqref{rmatrix} into \eqref{fundbivect} together with the
expressions \eqref{vecfields} for the left- and right invariant vector
fields, we obtain the Poisson brackets of these generating
functions $j_a$ and functions $F\in\cif(G^{n+2g})$, resulting in the
following alternative definition of the flower algebra.

\begin{definition} (Flower algebra for groups $\prgr$)
\label{flower}

The flower algebra $\gothf$ for gauge group $H=\prgr$ on a genus $g$
 surface $S_{g,n}$ with $n$ punctures is the commutative Poisson algebra
\bea
\label{flowerdef}
\gothf=S\left(\bigoplus_{k=1}^{n+2g}\gothg\right)\otimes\cif(G^{n+2g}),
\eea
where $S\left(\bigoplus_{k=1}^{n+2g}\gothg\right)$ is the symmetric envelope of
the real Lie algebra $\bigoplus_{k=1}^{n+2g}\gothg$, i.e. the polynomials with real
coefficients on the vector space $\bigoplus_{l=1}^{n+2g}\gothg^*$. In
terms of a fixed basis
$\mathcal{B}=\{j_a^\mi,j_a^{A_k},j_a^{B_k},\; $i=1,\ldots,n,\; k=1,\ldots,g,\;
a=1,\ldots,\text{dim}\,G\},  its Poisson structure is given by
\begin{align}
\label{poiss}
&\{j_a^X \otimes 1,j_b^X\otimes 1\}=-\flh abc j^X_c\otimes 1\nonumber\\
&\{j_a^X\otimes 1,j_b^Y\otimes 1\}=- \flh d b c j_c^Y\otimes(\delta_a^{\;\;d}-\Ad^*(u_X)_a^{\;\;d})\qquad\qquad\forall X,Y\in\{M_1,\ldots,B_g\},\, X<Y\nonumber\\
&\{j^\ai_a\otimes 1,j^\bi_b\otimes 1\}=-\flh a b c j^\bi_c\otimes 1\;\qquad\qquad\qquad\qquad\qquad\forall i=1,\ldots,g\nonumber\\
\nonumber\\
&\{j^\mi_a\otimes 1, 1\otimes F\}=-1\otimes (J_a^{R_\mi}+J_a^{L_\mi})F-1\otimes (\delta_a^{\;\;b}-\Ad^*(u_\mi)_a^{\;\;b})\left(\sum_{Y>\mi} (J_b^{R_Y}+J_b^{L_Y})F\right)\nonumber\\
&\{j^\ai_a\otimes 1, 1\otimes F\}=-1 \otimes (J^{R_\ai}_a+J^{L_\ai}_a)F-1\otimes (J_a^{R_\bi}+J_a^{L_\bi})F-1\otimes \Ad^*(u_\bi^\inv u_\ai)_a^{\;\;b}J_b^{R_\bi}\nonumber\\
&\qquad\qquad\qquad\qquad-1\otimes (\delta_a^{\;\;b}-\Ad^*(u_\ai)_{a}^{\;\;b})\left(\sum_{Y>\ai} (J_b^{R_Y}+J_b^{L_Y})F\right)\nonumber\\
&\{j^\bi_a\otimes 1, 1\otimes F\}=-1\otimes J_a^{L^\ai} F-1\otimes (J_a^{R_\bi}+J_a^{L_\bi})F\nonumber\\
&\qquad\qquad\qquad\qquad-1\otimes (\delta_a^{\;\;b}-\Ad^*(u_\bi)_a^{\;\;b})\left(\sum_{Y>\bi} (J_b^{R_Y}+J_b^{L_Y})F\right),
\end{align}
where $F\in\cif(G^{n+2g})$, $M_1<\ldots<M_n<A_1,B_1<\ldots<A_g,B_g$
and $J_a^{L_X}$, $J_a^{R_X}$ denote the right- and left invariant vector
fields \eqref{gvecfields} on the different copies of $G$.
\end{definition}


\subsection{The decoupling transformation}
\label{decoupsect}

We will now show how the flower algebra for  groups of type $H=\prgr$
 can be broken down into a set of Poisson commuting building
blocks. In doing this, we follow closely the work of Alekseev and
Malkin \cite{AMII} who treated the case of compact, semisimple Lie
 groups $H$. They give a bijective transformation
that maps the Poisson structure on the moduli space $\calm_{n,g}$ to
the direct sum of $n$ symplectic forms on $H$-conjugacy classes and
$g$ copies of the Heisenberg double $D_+(H)$. While, in general, this
transformation is quite complicated,  which makes it difficult to obtain
an explicit expression in terms of coordinates, the
picture is a lot simpler in the
case $H=\prgr$. Not only can the transformation of Alekseev and Malkin
be  generalised to this setting, but it is then possible to obtain an
explicit expression in terms of the generators defined in
Def.~\ref{flower}. This allows us to verify the asserted properties of
this transformation by direct calculation.

\pagebreak

\begin{definition}(Decoupling transformation)
\label{decoupdef}

The decoupling transformation is the bijective transformation
$K:\gothf\rightarrow\gothf$
\begin{align}
\label{decoup}
K:\qquad 1\otimes F\mapsto 1\otimes F,\;\;\; j_a^\mi\otimes 1\mapsto j_a^\mip,\;\;\;j_a^{A_j}\otimes 1\mapsto j_a^\ajp,\;\;\;j_a^{B_j} \otimes 1\mapsto j_a^\bjp
\end{align}
for $F\in\cif(G^{n+2g})$, $i=1,\ldots,n$ and $j=1,\ldots,g\nonumber$.
 The transformed generators are given by
\begin{align}
\label{transfj}
&j_a^{\mip}=j_a^\mi -
(\delta_a^{\;\;b}-\Ad^*(u_\mi)_a^{\;\;b})
\bigg(\sum_{k=i+1}^n\Ad^*
\big(u_{M_{k-1}}\cdots
 u_{M_{i+1}}\big)_b^{\;\;c}\cdot j_c^{M_k}\\
&\qquad\qquad\qquad\qquad\qquad\qquad\;+\sum_{k=1}^g\Ad^*\big(u_{K_{k-1}}\cdots
u_{K_1}\cdot u_{M_n}\cdots u_{M_{i+1}}\big)_b^{\;\;c}\cdot
j_c^{H_k}\bigg)\nonumber\\
&j_a^{\aip}=j_a^\ai+\Ad^*(u_\ai^\inv)_a^{\;\;b}(j_b^\ai-j_b^\bi)+(\Ad^*(u_\ai)-\Ad^*(u_\bi
u_\ai^{-1}))_a^{\;\;b}\bigg(\sum_{k=i+1}^g\Ad^*\big(u_{K_{k-1}}\cdots
u_{K_{i+1}}\big)_b^{\;\;c}j_c^{H_k}\bigg)\nonumber\\
&j_a^{\bip}=j_a^\bi+\Ad^*(u_\ai^\inv)_a^{\;\;b}(j_b^\ai-j_b^\bi)+(\Ad^*(u_\bi)-\Ad^*(u_\bi u_\ai^{-1}))_a^{\;\;b}\bigg(\sum_{k=i+1}^g\Ad^*\big(u_{K_{k-1}}\cdots u_{K_{i+1}}\big)_b^{\;\;c}j_c^{H_k}\bigg),\nonumber
\end{align}
where we write $j^X_a$ for $j^X_a\otimes 1$ and $F$ for $1\otimes F$. The expressions
$\Ad^*(u_X)_a^{\;\;b}:\;(u_{M_1},\ldots,u_{B_g})\mapsto
\Ad^*(u_X)_a^{\;\;b}$ for $X\in\{M_1,\ldots,B_g\}$ are to be
interpreted as functions in $\cif(G^{n+2g})$, and we set
\begin{align}
\label{jothi}
&u_{K_i}:=u_\bi u_\ai^{-1}u_\bi^{-1}u_\ai\\
&j_a^{H_i}:=\big(\delta_a^{\;\;b}-\Ad^*(u_\ai^{-1}u_\bi^\inv u_\ai)_a^{\;\;b}\big)j_b^\ai+\big(\Ad^*(u_\ai^{-1}u_\bi^\inv
u_\ai)-\Ad^*(u_\bi^{-1} u_\ai)\big)_a^{\;\;b}j_b^\bi.\nonumber
\end{align}
The  inverse of $K$  is given by
\begin{align}
\label{invdecoup}
&j_a^\mi=j^\mip_a+(\delta_a^{\;\;b}-\Ad^*(u_\mi)_a^{\;\;b})\left(\sum_{k=i+1}^n
  j_b^{M'_k}\right)
\\
&\qquad+(\delta_a^{\;\;b}-\Ad^*(u_\mi)_a^{\;\;b})\left(\sum_{k=1}^g
\left(\delta_b^{\;\;c}-\Ad^*({u_{A_k}^\inv})_b^{\;\;c}\right)j_c^{A'_k}+\left(\Ad^*({u_{A_k}^\inv})-\Ad^*({u^{-1}_{B_k}
  u_{A_k}})\right)_b^{\;\;c}j_c^{B'_k}\right)\nonumber\\
&j_a^\ai=j_a^\aip-\Ad^*(u_\ai^\inv)_a^{\;\;b}(j_b^\aip-j_b^\bip)\nonumber\\
&\qquad+(\delta_a^{\;\;b}-\Ad^*(u_\ai)_a^{\;\;b})\left(\sum_{k=i+1}^g
\left(\delta_b^{\;\;c}-\Ad^*({u_{A_k}^\inv})_b^{\;\;c}\right)j_c^{A'_k}+\left(\Ad^*({u_{A_k}^\inv})-\Ad^*({u^{-1}_{B_k}
  u_{A_k}})\right)_b^{\;\;c}j_c^{B'_k}\right)\nonumber\\
&j_a^\bi=j_a^\bip-\Ad^*(u_\ai^\inv)_a^{\;\;b}(j_b^\aip-j_b^\bip)\nonumber\\
&\qquad+(\delta_a^{\;\;b}-\Ad^*(u_\bi)_a^{\;\;b})\left(\sum_{k=i+1}^g
\left(\delta_b^{\;\;c}-\Ad^*({u_{A_k}^\inv})_b^{\;\;c}\right)j_c^{A'_k}+\left(\Ad^*({u_{A_k}^\inv})-\Ad^*({u^{-1}_{B_k}
  u_{A_k}})\right)_b^{\;\;c}j_c^{B'_k}\right)\nonumber.
\end{align}
\end{definition}

With this definition, we can calculate the transformed bracket and verify that the transformation does indeed decouple the mixed contributions in \eqref{poiss} into Poisson-commuting building blocks.
\begin{theorem}
\label{decth}
The decoupling transformation $K$ maps the Poisson structure
\eqref{poiss} to the direct sum of $n$ Poisson structures on the dual
$H^*$  and $g$ copies of the symplectic structure of the Heisenberg
double $D_+(H)$:
\begin{align}
\label{partbr}
&\{j_a^\mip,j_b^\mjp\}=-\delta_{ij}\flh a b c j^\mip_c\\
&\{j_a^\mip, 1\otimes F\}=-1\otimes (J_a^{R_\mi}+J_a^{L_\mi})F\qquad\forall i,j=1,\ldots,n\nonumber\\
\nonumber\\
\label{hbr}
&\{j_a^\aip,j_b^\ajp\}=-\delta_{ij}\flh a b c j^\aip_c\\
&\{j_a^\bip,j_b^\bjp\}=-\delta_{ij}\flh a b c j^\bip_c\nonumber\\
&\{j_a^\aip,j_b^\bjp\}=-\delta_{ij}\flh a b c j^\bip_c\nonumber\\
&\{j_a^\aip,1\otimes F\}=-1\otimes J_a^{R_\ai} F-1\otimes (1+\Ad^*(u_\bi^\inv u_\ai))_a^{\;\;b}J_b^{R_\bi} F \nonumber\\
&\{j_a^\bip,1\otimes F\}=-1\otimes J_a^{R_\bi} F\qquad\qquad\qquad\qquad\qquad\qquad\qquad\qquad\forall i,j=1,\ldots,g\nonumber\\
\nonumber\\
&\{j_a^\mip,j_b^\ajp\}=\{j_a^\mip,j_b^\bjp\}=0\qquad\forall i=1,\ldots,n,\,j=1,\ldots,g.
\end{align}
\end{theorem}

{\bf Proof:} The Poisson brackets \eqref{partbr}, \eqref{hbr} of the
 transformed generators $j^\mip_a,j^\aip_a,j^\bip_a$ can be calculated
 directly from the Poisson brackets \eqref{poiss} of the flower
 algebra. We then insert the expressions
 \eqref{vecfields} for the vector fields on $H$ in the Poisson
 bivectors \eqref{conjbivect} and \eqref{heisdouble} of the dual $H^*$
 and the Heisenberg double $D_+(H)$ and apply them to the functions
 $j_a\in\cif(H): (u,-\Ad^*(u^\inv)\bj)\mapsto j_a$ to verify that the
 result does agree with  the decoupled brackets \eqref{partbr} and
 \eqref{hbr}.$\hfill\Box$

Note that $K$ may also be viewed as the pullback of a map $H^{n+2g}\rightarrow
H^{n+2g}$  which is the identity on $G^{n+2g}$ and leaves each of the conjugacy
classes $\calc_{\mu_i s_i}$ invariant.
From
\eqref{transfj} we see that this map
 adds to $\bj_\mi$  respectively $\bj_\mip$ an  element of
$\bigoplus_{k=1}^{n+2g}\gothg\otimes\cif(G^{n+2g})$ preceded by a factor
$(1-\Ad^*(u_\mi))$. It follows from \eqref{conjparam} that such
transformations map a given conjugacy class into itself.

The transformation $K$ simplifies the Poisson structure of the flower
algebra considerably by decoupling the contributions of
different punctures and handles. However, it is still possible to
simplify the resulting Poisson structure further by breaking up the
Heisenberg double Poisson structure \eqref{hbr} associated to each handle.
With the transformation  $L:H^{2g+n} \rightarrow H^{2g+n}$
\bea
\label{hdecoup}
(u_\ai,u_\bi) &\mapsto   &(w_{1,i},w_{2,i})=(u_\ai,u_\bi^\inv u_\ai)
\\
(\bj_\aip,\bj_\bip)  &\mapsto
& (\bk_{1,i},\bk_{2,i})=(\bj_\aip-\bj_\bip,\bj_\bip)\qquad
i=1,\ldots,g\nonumber\\
u_{M_i}& \mapsto & u_{M_i}\nonumber\\
\bj_{M'_i} & \mapsto &\bj_{M'_i} \qquad\qquad\quad
\qquad\qquad\qquad\qquad i=1,\ldots,n,\nonumber
\eea
we map each Heisenberg double into the
 cotangent bundle symplectic
structure $T^*(G\times G)$:
\begin{align}
\label{hcotan}
&\{k_a^{1, i},k_b^{1, i}\}=-\flh a b c k_c^{1,i}\qquad\qquad\{k_a^{2, i},k_b^{2, i}\}=-\flh a b c k_c^{2,i}\qquad\qquad\{k^a_{1, i},k^b_{2, i}\}=0\\
&\{k_a^{1,i},F\}(w_{1,i},w_{2,i})=-\frac{\text{d}}{\text{d}t}|_{t=0}F(w_{1,i}\, e^{tJ_a}\,,\,
w_{2,i})\nonumber\\
&\{k_a^{2,i},F\}(w_{1,i},w_{2,i})=-\frac{\text{d}}{\text{d}t}|_{t=0}F(w_{1,i}\,,\,e^{-tJ_a}w_{2,i}).\nonumber
\end{align}
Combining the decoupling transformation $K$ with the pull-back  $L^*$
 and using the notation \eqref{conjmap}  then yields the following theorem
\begin{theorem}
\label{compldec}
The bijective map $L^*\circ K:
\gothf\rightarrow \gothf$  maps the Poisson
structure \eqref{poiss} of the flower algebra  to the direct sum of
$n$ copies of the dual Poisson-Lie group $H^*$ and  $2g$
copies of the cotangent bundle Poisson structure $T^*G$.
The symplectic leaves of the Poisson manifold
$\big(H^*\big)^n\times\big(T^*G\big)^{2g}$
are of the form
 $\calc_{\mu_1 s_1}\times\ldots\times\calc_{\mu_n s_n}\times
H^{2g}$ and the  pull-back of the symplectic structure on each
leaf via the map
\bea
\mbox{conj}_{\mu_1 s_1}\times \ldots \times \mbox{conj}_{\mu_n s_n}
\times \mbox{id}^{2g}:
H^{n+2g}& \rightarrow& \calc_{\mu_1 s_1}\times\ldots\times\calc_{\mu_n
  s_n}\times
H^{2g}
\eea
is the exterior derivative of the symplectic potential
\bea
\label{symppot}
\Theta=-\sum_{i=1}^n \langle dv_\mi v_\mi^\inv\,,\, j^\mip_a
P^a\rangle+\sum_{i=1}^g
\langle w_{1,i}^\inv dw_{1,i}\,,\,k_a^{1,i}P^a\rangle
 -\langle dw_{2,i}w_{2,i}^\inv\,,\, k_a^{2,i}P^a\rangle.
\eea
\end{theorem}
{\bf Proof:} The expression for the pull-back of
the  symplectic potential on the
symplectic leaves of $H^*$ was derived for the case
$G=\widetilde{SO(2,1)}$ in \cite{we}, but the derivation is
valid  for any Lie group $G$. The expression for the symplectic
potential on $T^*(G\times G)$ is standard and uses the identification
of $T^*G$ with $G\times \gothg^*$. The identification can be
made using either the left- or the right-multiplication of $G$;  our
definition  of $k_a^{1,i}$ and $k_a^{2,i}$ is such  that we use
right-multiplication for one copy and left-multiplication for
the other copy of $G$ in $T^*(G\times G)$.
The  reasons for this choice are related to the natural
action of the quantum double of $G$ in the quantum theory, and
will become clear in Sect.~\ref{qudouble}. $\hfill\Box$

This theorem provides us with an interpretation of the map
$L^*\circ K$.
 In the decoupled coordinates
$\bj_\mip,u_\mip, \bk_{\alpha,i}$ and $w_{\alpha,i}$ the symplectic
structure on symplectic leaves has the canonical form \eqref{symppot}.
We shall see in Sect.~3  that the
Poisson structure expressed in terms of the decoupled coordinates
is amenable to a rather straightforward quantisation procedure.
The map $L^*\circ K$ thus establishes a link between
two important sets of coordinates, the holonomy coordinates with
a direct gauge-theoretical
 interpretation and the decoupled coordinates
which are convenient for quantisation.

\subsection{Symmetries}

As the flower algebra for the group $H=\prgr$ is closely related to
 the phase space of Chern-Simons gauge theory with gauge group $H$, the
 moduli space of flat $H$-connections, it is to be expected that at least
 some of the invariance transformations of the underlying Chern-Simons
 theory give rise to symmetries of the flower algebra. We identify two such groups of symmetries of the flower algebra
 and show how they are related to the invariance transformations of
 the underlying Chern-Simons gauge theory.

 As a topological field theory, Chern-Simons
theory on a surface $S_{g,n}$ is invariant under diffeomorphisms of this
surface, in particular, under large diffeomorphisms, which form the
mapping class group $\text{Map}(S_{g,n})$. In \cite{we} it was shown
for the case $H=\widetilde{SO(2,1)}\ltimes \RR^3$
that the mapping
class group $\text{Map}(S_{g,n}\!\!\setminus \!D)$ of the surface with
a disc removed acts on the flower algebra as  Poisson isomorphisms.
The proof
can be extended   to general $H$, but  we will not give it here.
Instead we defer a full discussion of the mapping class group
in classical and in particular quantised Chern-Simons theories
with gauge groups $G\ltimes \gothg^*$ to a separate  paper \cite{we3}.

The second type  of symmetry acting on the flower algebra is related
to the other class of invariance transformations in Chern-Simons
theory,
Chern-Simons
gauge transformations.
 In the description of Chern-Simons gauge theory by means of a set of
 curves representing the generators of the fundamental group, Chern-Simons gauge transformations that are
 nontrivial at the basepoint act on the associated holonomies by
 global conjugation with an element of $H$. Via the
 identification of the holonomies with the different copies of $H$ in
 definitions Def.~\ref{flower0}, Def.~\ref{flower}, this action on the
 holonomies induces transformations of the flower algebra. However,
 these transformations are in general {\em not} Poisson
 isomorphisms unless we take into account the nontrivial Poisson
 structure \eqref{grbivect} of the group $H$. For semisimple $H$, it
 was shown in
 \cite{FR}
that, interpreted as
 maps $H\times H^{n+2g}\rightarrow H^{n+2g}$, they are Poisson
 isomorphisms with respect to the flower algebra Poisson structure on
 $H^{n+2g}$ and the Poisson structure on $H\times H^{n+2g}$ that is
 the direct product of the Poisson structure \eqref{grbivect} on $H$
 and the flower algebra Poisson structure on $H^{n+2g}$. However, we will see that there
is a much larger group of Poisson symmetries of the flower
 algebra that generalises the conjugation with elements of the
group $H$. In particular, for the case where the
 exponential map $\exp:\gothg\rightarrow G$ is bijective, these
 Poisson symmetries give rise to a Poisson action of $H$ on the flower
 algebra that can be interpreted as a deformed  conjugation.

\begin{theorem}(Action of $G\ltimes\cif(G)$)
\label{defconjact}
\vspace{-0.3cm}
\begin{enumerate}
\item  Consider the group $G\ltimes\cif(G)$ with multiplication  law
\bea
\label{defmult}
(h_1, f_1 )\cdot(h_2,f_2)=(h_1 h_2, f_1+f_2\circ\Ad_{h_1^\inv})\quad\forall h_1,h_2\in G,\,f_1,f_2\in\cif(G).
\eea
It acts on the flower algebra via
\begin{align}
\label{actiondef}
(h,f):\;\; &j^X_a\otimes 1 & &\mapsto & &j_b^X\otimes \Ad^*(h)_a^{\;\;b}+1\otimes(\Ad^*(h)_a^{\;\,b}\{j^X_b\otimes 1,1\otimes f\circ\Phi_\infty\}) \\
& 1\otimes F & &\mapsto & &1\otimes F\circ Ad_{h^\inv}^{n+2g},\nonumber
\end{align}
where  $\Ad_{h}^{n+2g}:\; (u_{M_1},\ldots,u_{B_g})\mapsto( h
u_{M_1}h^\inv,\ldots,h u_{B_g}h^\inv)$ denotes the global conjugation
by $h\in G$ , $\{\;,\;\}$ the Poisson bracket \eqref{poiss} on the
flower algebra and $\Phi_\infty: G^{n+2g}\rightarrow G$ is the map
\bea
\label{utot}
\Phi_\infty(u_{M_1},\ldots,u_{B_g})=u_{tot}=u_{K_g}\cdots u_{K_1}\cdot u_{M_n}\cdots u_{M_1}
\eea
with $u_{K_i}$ given by \eqref{jothi}.
\vspace{-0.1cm}
\item This  action is a Poisson action. The infinitesimal generators
  of the action of $G\subset G\ltimes\cif(G)$ are the elements
\begin{align}
\label{jtot}
j^{tot}_a=&\sum_{i=1}^n j_b^\mi\otimes \Ad^*(u_{M_{i-1}}\cdots
u_{M_1})_a^{\;\;b}+\sum_{i=1}^g 1\otimes \Ad^*( u_{K_{i-1}}\cdots
u_{M_1})_a^{\;\;b} \cdot j_b^{H_i}\\
=&\sum_{i=1}^n j_a^\mip+\sum_{k=1}^g 1\otimes (\delta_a^{\;\;b}- \Ad^*(u_{A_k}^\inv)_a^{\;\;b})\cdot j_b^{A'_k}+1\otimes (\Ad^*(u_{A_k}^\inv)-\Ad^*(u_{B_k}^\inv u_{A_k}))_a^{\;\;b}\cdot j_b^{B'_k},\nonumber
\end{align}
with $j_a^{H_i}$ given by \eqref{jothi}, and
the action of $\cif(G)\subset G\ltimes\cif(G)$ is generated by
 functions $1\otimes (f\circ\Phi_\infty)$, $f\in\cif(G)$. We have for
 all elements $\varphi\in\gothf$
\begin{align}
\label{infgen}
&\frac{\text{d}}{\text{d}t}|_{t=0}(e^{-tJ_a}\,,\,0)(\varphi)=\{j_a^{tot},\varphi\}
& &\frac{\text{d}}{\text{d}t}|_{t=0}(1\,,\,-t\cdot f)(\varphi)=\{1\otimes (f\circ\Phi_\infty)\, ,\,\varphi\}.
\end{align}
\end{enumerate}
\end{theorem}
{\bf Proof:}
That \eqref{actiondef} defines a group action of $G\ltimes\cif(G)$ on the
flower algebra with infinitesimal generators \eqref{infgen} can be
shown by direct computation using the Poisson brackets \eqref{poiss}
or, alternatively, \eqref{partbr},\eqref{hbr} of the flower
algebra. That it is a Poisson action follows from the fact that it is
infinitesimally generated via the Poisson brackets \cite{MaRa},  but can also be verified directly from \eqref{actiondef} and the Poisson bracket of the flower algebra. $\hfill\Box$

Note that the map $\Phi_\infty: (u_{M_1},\ldots, u_{B_g})\mapsto
u_{tot}$ as well as the algebra elements $j_a^{tot}$ occurring in
Theorem \ref{defconjact} have a geometric meaning \cite{we}. If we
parametrise the holonomies $M_1,\ldots,B_g$ associated to
the generators $m_1,\ldots,b_g$ of $\pi_1(S_{g,n})$
according to \eqref{gparam}
\bea
(u_X,-\Ad^*(u_X^\inv)\bj_X)=(u_X,-\Ad^*(u_X^\inv)j^X_a P^a)\qquad\forall X\in\{M_1,\ldots,B_g\}
\eea and identify the parameters $j_a^X$ with generators $j^X_a\otimes 1$, then the holonomy associated to the curve $k_\infty$ defined in \eqref{pirel} is
$\text{Hol}(k_\infty)=(u_{tot}\,,\,-\Ad^*(u_{tot}^{-1})\bj_{tot})$
with $u_{tot}$ and $\bj_{tot}$ given by \eqref{utot} and \eqref{jtot}.

We will now relate this action of the group $G\ltimes\cif(G)$ on the
flower algebra to the transformations
 induced by simultaneous conjugation of the holonomies with a fixed
 element of $\prgr$. From the group multiplication law
 \eqref{groupmult}, it follows that $(h,\bx)\in\prgr$ acts on elements
of the flower algebra by conjugation with  the inverse
 $(h^\inv, -\Ad^*(h)\bx)$ according to
\begin{align}
\label{obsconj}
&1\otimes F & &\mapsto & &1\otimes (F\circ\Ad_{h^\inv}^{n+2g})\\
 &j^X_a\otimes 1 & &\mapsto & &j_b^X\otimes \Ad^*(h)_a^{\;\;b}-1\otimes(\Ad^*(h)-\Ad^*(u_Xh))_a^{\;\;b}x_b.\nonumber
\end{align}
This action is not a Poisson action.
On the other hand, we can use \eqref{poiss} together with the
definition of the map $\Phi_\infty$ to evaluate the bracket
$\{j^X_a\,,\,f\circ\Phi_\infty\}$ in the definition \eqref{actiondef}
of the $G\ltimes \cif(G)$-action on the flower algebra and obtain
the Poisson action
\begin{align}
\label{conjact}
(h,f):\;\; &1\otimes F & &\mapsto & &1\otimes (F\circ\Ad_{h^\inv}^{n+2g})\\
 &j^X_a\otimes 1 & &\mapsto & &j_b^X\otimes \Ad^*(h)_a^{\;\;b}-1\otimes((\Ad^*(h)-\Ad^*(u_Xh))_a^{\;\;b}(J_b^Lf)\circ\Phi_\infty).\nonumber
\end{align}
Comparing expressions \eqref{obsconj} and \eqref{conjact}, we see
that the action of $G\ltimes\cif(G)$ on the flower algebra can be
interpreted as a generalised conjugation: the action of $G\subset
G\ltimes\cif(G)$ agrees with the action of $G$ via global conjugation,
whereas the action of $\cif(G)\subset G\ltimes\cif(G)$ mimics global
conjugation with $\gothg^*$, only that now the
transformation vector $\bx\in\gothg^*$ is replaced
by a function of the group element $u_{tot}$.

If the exponential map $\exp:\gothg\rightarrow G$ is bijective, group
elements $u\in G$ can be parametrised as $u=\exp(p^a J_a)$ and the
relation between the action of $G\ltimes\cif(G)$ and the action of the
group $\prgr$ by global conjugation becomes more explicit. Denoting the
inverse of $\exp:\gothg\rightarrow G$ by $\log: G\rightarrow\gothg$,
we can define an embedding $\iota:\gothg^*\rightarrow \cif(G)$ via
\bea
\label{iotadef}
\iota(\bq)(u)=\langle \bq,\log(u)\rangle\qquad \forall
\bq\in\gothg^*,u\in G,
\eea
where $\langle\,,\,\rangle$ denotes the pairing \eqref{blf}. In
particular, we have the coordinate functions $\iota(P^a):
u=\exp(p^bJ_b)\mapsto p^a$. Then, the group
multiplication \eqref{defmult} restricted to
$G\ltimes \iota(\gothg^*)$ becomes simply the group
multiplication of $H$, and we obtain the following lemma
\begin{lemma}
\label{symmact}
If the exponential map $\exp:\gothg\rightarrow G$ is bijective, there
is a Poisson action of the group $\prgr$ on the flower algebra given
by
\begin{align}
\label{poisact}
(h,\bq):\;\;  &1\otimes F & &\mapsto &  &
1\otimes (F\circ\Ad_{h^\inv}^{n+2g})\\
&  j^X_a\otimes 1 &  &\mapsto & &j_b^X\otimes
\Ad^*(h)_a^{\;\;b}-1\otimes
(\Ad^*(h)-\Ad^*(u_Xh))_a^{\;\;b}x_b.\nonumber
\end{align}
 where $q_b= T_b^{\;\;c}(u_{tot})\,x_c$
and
\bea
T(u_{tot})= \frac{1 - \Ad^*(u_{tot})}{\ad^*(p^a_{tot}J_a)}
\eea
is a linear map depending on the total holonomy $u_{tot}$.
\end{lemma}

{\bf Proof}:  Apply the formula \eqref{conjact} for the Poisson action
of $G\ltimes\cif(G)$  to the function $f=\iota(\bq)$.
The lemma then follows from  the formula $(J_b^L \iota(\bq))\circ\Phi_\infty = \left(T^{-1}\right)_b^{\;\;c}(u_{tot})\,q_c$,
which can be found, for example, in \cite{AMaP}, p. 179.
 $\hfill\Box$

\section{Quantisation}
\label{quantisation}

By Theorem \ref{compldec}, we have reduced the task of quantising the
flower algebra to the quantisation of the symplectic structure on the
cotangent bundle $T^*G$ of the group $G$ and the dual Poisson structure
 $H^*$. Both of these Poisson structures are relatively simple and  special cases of the following general situation.
We have a commutative Poisson algebra given as a tensor product
\begin{align}
\label{poissalgdef}
\calq=S(\gothe)\otimes\cif(E),
\end{align}
where $E$ is a finite dimensional, simply connected Lie group with
Lie algebra $\gothe=\text{Lie}\,E$ and $S(\gothe)$ denotes the
symmetric
envelope of $\gothe$. The Poisson structure is that of a semidirect
product:
Poisson brackets of two elements of $S(\gothe)$ are given by the
derivative extension of the Lie bracket on $\gothe$, Poisson brackets
of two functions in $\cif(E)$ vanish and Poisson brackets of elements
of $S(\gothe)$ with functions in $\cif(E)$ are derived from a group
action $.$ of $E$ on itself
\begin{align}
\label{genpb}
&\{\xi\otimes 1,\eta\otimes 1\}=[\xi,\eta]_\gothe\otimes 1\qquad\qquad
\{1\otimes F,1\otimes K\}=0\\
&\{\xi\otimes 1, 1\otimes F\}(d)=\frac{\text{d}}{\text{d}t}|_{t=0}
F( e^{-t\xi}\,.u\,) \qquad\qquad\qquad\forall
\xi,\eta\in\gothe,\;u\in E,\;F,K\in\cif(E).\nonumber
\end{align}
 In the case of the
cotangent bundle $T^*G$, we have $E=G$ and the group action is the
inverse left multiplication or, alternatively, right multiplication by
$G$. For the dual Poisson structure $H^*$, it is conjugation with
$G$. We proceed now to discuss the quantisation of a general
Poisson algebra $\calq=S(\gothe)\otimes\cif(E)$ of this type and then
apply the results to the cotangent bundle $T^*G$, the dual $H^*$ and
the decoupled flower algebra in Sect.~\ref{decoupquantsect}.

\subsection{Quantisation of Poisson algebras $\calq=S(\gothe)\otimes\cif(E)$}
\label{sdirquant}

Let  $\calq=S(\gothe)\otimes\cif(E)$ be a Poisson algebra with a
semidirect product Poisson structure arising
from an action $\,.\,$ of the simply connected Lie group $E$ on itself as in \eqref{genpb}.
The Poisson algebra $\calq$ inherits a $\mathbb{N}$-grading from the canonical $\mathbb{N}$-grading of the symmetric envelope
\begin{align}
\label{gengrad}
\calq=\bigoplus_{k=0}^\infty \calq^{(k)}\qquad \calq^{(k)}=
S^{(k)}(\gothe)\otimes\cif(E),
\end{align}
 where $S^{(k)}(\gothe)$ is the space of polynomials
of degree $k$ in a basis $\mathcal{B}_\gothe=\{\xi_1,\ldots, \xi_{\text{dim}\,E}\}$ with real coefficients. The multiplication of homogeneous elements adds their degrees, whereas the Poisson bracket \eqref{genpb} adds their degrees and subtracts one
\begin{align}
\label{gengrad2}
&\calq^{(k)}\cdot\calq^{(l)}\subset\calq^{(k+l)} & &\{\calq^{(k)},\calq^{(l)}\}\subset\calq^{(k+l-1)}.
\end{align}

In quantisation, the commutative Poisson algebra $\calq$ is to be
replaced
by an associative *-algebra $\calqq$, which depends on a deformation
parameter
$\hbar$ and  has to exhibit certain structural properties relating it
to
the classical algebra $\calq$. Every element of the quantum algebra
$\calqq$
must correspond to a unique element  in the complexified classical
algebra $\calq_\CC$
and conversely, it must be possible to assign to every element of
the complexified classical algebra
 a quantum counterpart in $\calqq$. This is equivalent to the
 existence of
a vector space isomorphism $Q:\calq_\CC\rightarrow \calqq$. In order to
obtain an algebra $\calqq$ that merits the name quantisation, we must
furthermore request that the product of two of its elements  is
given as the quantum counterpart of the product of the corresponding
classical elements  plus a quantum correction of order $O(\hbar)$,
and that their commutator is equal to $i\hbar$ times the quantum
counterpart of the Poisson bracket of the corresponding classical
elements  plus a quantum correction of order $O(\hbar^2)$:
\begin{align}
\label{quantcond}
&Q(W)\cdot Q(Z)=Q(W\cdot Z)+O(\hbar)\\
\label{quantcond2}
&[Q(W),Q(Z)]=i\hbar Q(\{W,Z\})+O(\hbar^2)\qquad\forall W,Z\in\calq.
\end{align}
In general, the quantum corrections
in \eqref{quantcond2} cannot be eliminated for all elements $W,Z\in\calq$. 
(For the case of the Heisenberg algebra, this is a consequence of the no-go theorem by Groenewold and van Hove \cite{Groenewold,VanHove1,VanHove2}.)
But there should be a Poisson subalgebra of the classical algebra
$\calq$ containing the generating elements $\xi_1\otimes1,
\ldots,\xi_{\text{dim}\,E}\otimes 1$ and $1\otimes F$,
for which this is possible.

As our Poisson algebra $\calq$ is of a particularly simple type and
related to the symmetric envelope of a Lie algebra $\gothe$,
a framework for the construction of the quantum algebra is provided
by the theory of universal enveloping algebras and the theorem
of Poincar\'e-Birkhoff-Witt \cite{dixmier}. We obtain the
following theorem defining a quantum algebra with the requested properties
\begin{theorem}
(Construction of the quantum algebra $\calqq$)
\label{genqao}

Let $\calqq$ be the associative algebra
$\calqq=U(\gothe)\tenltimes\cif(E,\mathbb{C})$
with a semidirect multiplication defined by
\bea
\label{genmult}
(\xi\otimes F)\cdot(\eta\otimes K)=\xi\cdot_U\eta\otimes FK+i\hbar\;
\eta
\otimes F\, \{\xi\otimes 1,1\otimes K\}\quad\forall \xi,\eta\in\gothe,\, F,K\in\cif(E,\mathbb{C}),
\eea
where $U(\gothe)$ denotes the universal enveloping algebra of the Lie
algebra $\gothe$ with Lie bracket multiplied by a factor $i\hbar$,
 $\cdot_U$ the multiplication in $U(\gothe)$ and $\{\,,\,\}$ the Poisson bracket \eqref{genpb}. Then
\begin{enumerate}
\item $\widehat{\calq}$ has a *-structure given by $(\xi\otimes
  1)^*=\xi\otimes 1$, $(1\otimes F)^*=1\otimes \bar F$ for $\xi\in \gothe$,
  $F\in\cif(E,\CC)$.
\item The algebra $\calqq$ inherits a filtration from the canonical filtration of the universal enveloping algebra $U(\gothe)$
\begin{align}
\label{genfilter1}
&\calqq=\bigcup_{k=0}^\infty \calqq^{(k)}\qquad \calqq^{(k)}=U^{(k)}(\gothe)\otimes\cif(E,\mathbb{C})\subset\calqq^{(k+1)}\qquad\forall k\in\mathbb{N}
\end{align}
with $\calqq^{(0)}\cong\cif(E,\mathbb{C})\cong\calq^{(0)}_\CC$, $\calqq^{(k)}/\calqq^{(k-1)}\cong\calq^{(k)}_\CC$ $\forall k\geq 1$ and
\bea
\label{multpoiss}
\calqq^{(k)}\cdot\calqq^{(l)}\subset\calqq^{(k+l)}\qquad[\calqq^{(k)},\calqq^{(l)}]\subset\calqq^{(k+l-1)}\qquad\forall k,l\in\mathbb{N}.
\eea
In particular, the commutator $[\,,\,]$ of $\calqq$ defines a Lie bracket on the space $\calqq^{(1)}$.
\item In terms of elements $\xi_1<\ldots<\xi_{\text{dim}\,E}$ of an ordered basis of $\gothe$, a vector space isomorphism $Q:\calq_\CC\rightarrow \calqq$ is defined by
\begin{align}
&\label{quantmap}
Q(\xi_{a_1}\ldots\xi_{a_m}\otimes F)
:=\xi_{a_1}\cdot_U\ldots \cdot_U\xi_{a_m}\otimes F=Q(1\otimes F)\cdot Q(\xi_{a_1}\otimes 1)\cdots Q(\xi_{a_m}\otimes 1)\nonumber\\
&\forall a_1\leq\ldots\leq a_m, F\in\cif(E,\mathbb{C}).
\end{align}
It satisfies
\begin{align}
\label{genqucond1}
&\Pi^{(k+l)}\big(\,Q(W)\cdot Q(Z)-Q(WZ)\,\big)=0\\
\label{genqucond2}
&\Pi^{(k+l-1)}\big(\,[Q(W), Q(Z)]-i\hbar\;Q(\{W,Z\})\,\big)=0\qquad\forall W\in\calq^{(k)}_\CC, Z\in\calq^{(l)}_\CC,
\end{align}
where $\Pi^{(k)}$ is the canonical projection $\calqq^{(k)}\rightarrow
 \calqq^{(k)}/\calqq^{(k-1)}$.
In particular,
$Q|_{(\calq^{(0)}\oplus\calq^{(1)})_\CC}:\,\calq^{(0)}_\CC\oplus\calq^{(1)}_\CC
\rightarrow \calqq^{(1)}$ is a Lie algebra isomorphism with
respect to the brackets $\{\,,\,\}|_{(\calq^{(0)}\oplus\calq^{(1)})_\CC }$
and  $[\,,\,]|_{\calqq^{(1)}}$.
\end{enumerate}
\end{theorem}
Note that for general elements $\theta, \chi\in U(\gothe)$ of the
 universal enveloping algebra $U(\gothe)$ the multiplication law defined by
 \eqref{genmult} can be written using  Sweedler notation
\bea
\label{sweedlermult}
(\theta\otimes F)\cdot(\chi\otimes K)=\sum (\theta_{(1)}\cdot_U \chi)\otimes(F\cdot
\theta_{(2)}^L K),
\eea
where $\Delta_U: U(\gothe)\rightarrow U(\gothe)\otimes U(\gothe)$,
$\Delta_U(\theta)=\sum
  \theta_{(1)}\otimes \theta_{(2)}$ is the co-multiplication of the universal
 enveloping algebra defined inductively
by $\Delta_U(\xi)=1\otimes \xi+\xi\otimes
 1$ for $\xi\in\gothe$ and $\Delta_U(\theta\cdot_U
 \chi)=\Delta_U(\theta)\cdot_U\Delta_U(\chi)$ for $\theta,\chi\in U(\gothe)$.

{\bf Proof:}
1. The canonical filtration of universal enveloping
algebra $U(\gothe)=\bigcup_{k=0}^\infty U^{(k)}(\gothe)$ ,
where $U^{(k)}(\gothe)$ is the space of non-commutative polynomials
of degree $\leq k$ in the generators of the Lie algebra $\gothe$, satisfies
\begin{align}
\label{envmultpoiss}
&U^{(k)}(\gothe)\cdot_U U^{(l)}(\gothe)\subset U^{(k+l)}(\gothe),\;\;
[U^{(k)}(\gothe),U^{(l)}(\gothe)]_U\subset
U^{(k+l-1)}(\gothe)\;\forall k,l\in
\mathbb{N}
\end{align}
and $U^{(k)}(\gothe)/U^{(k-1)}(\gothe)\cong S^{(k)}(\gothe)$ for
 $k\geq 1$.
 This implies $\calqq^{(k)}/\calqq^{(k-1)}=(U^{(k)}(\gothe)\otimes
\cif(E,\mathbb{C}))/(U^{(k-1)}(\gothe)\otimes\cif(E,\mathbb{C}))\cong
 (U^{(k)}(\gothe)/U^{(k-1)}
(\gothe))\otimes\cif(E,\mathbb{C})\cong
 S^{(k)}(\gothe)\otimes\cif(E,\mathbb{C})=\calq^{(k)}_\CC$
for $k\geq1$. The identities \eqref{multpoiss} then follow directly
from \eqref{envmultpoiss} and the multiplication law \eqref{genmult}.
In particular, $[\calqq^{(1)},\calqq^{(1)}]\subset\calqq^{(1)}$,
which makes the subspace $\calqq^{(1)}$ with the commutator a Lie algebra.

\noindent 2. According to the theorem of Poincar\'e-Birkhoff-Witt, see for
  example
\cite{dixmier}, an ordered basis of the vector space $U(\gothe)$ is
given by the ordered monomials $\xi_{a_1}\cdot\ldots\cdot\xi_{a_m}$,
$a_1\leq\ldots\leq a_m$, $m\in\mathbb{N}$, in the elements of an ordered
  basis
$\xi_{1}<\ldots<\xi_{\text{dim}\,E}$ of the Lie algebra $\gothe$.
Therefore, the map $Q$ in \eqref{quantmap} defines a vector space
isomorphism from $\calq_\CC$ to $\calqq$. From the multiplication law
  \eqref{genmult} and the definition \eqref{genpb} of the Poisson
  bracket we have for the commutator of two Lie algebra elements
  $\xi,\eta\in\gothe$ and the commutator of a Lie algebra element with a function $F\in\cif(E,\mathbb{C})$
\begin{align}
\label{gencomm}
&[\xi\otimes 1,\eta\otimes
  1]=(\xi\cdot_U\eta-\eta\cdot_U\xi)\otimes 1= i\hbar \{\xi\otimes
  1,\eta\otimes 1\}\\
 &[\xi\otimes 1,
  1\otimes F]=i\hbar\, 1\otimes \left(-\frac{d}{dt}|_{t=0} F(e^{-t\xi}.\,\cdot)\right)=i\hbar\{\xi\otimes 1,1\otimes
  F\}.\nonumber
\end{align}
For elements $W\in\calq^{(k)}_\CC$, $Z\in\calq^{(l)}_\CC$ , the  product
  $Q(W)\cdot Q(Z)$
differs from $Q(WZ)$ only by factor ordering, which can be seen from
  the  multiplication law \eqref{genmult} and commutators
  \eqref{gencomm} to give rise to a
quantum correction in $\widehat{\calq}^{(k+l-1)}$ preceded by a factor $\hbar$.
The same applies to the commutator $[Q(W),Q(Z)]$ and $i\hbar
  Q(\{W,Z\})$,
only that now the quantum correction is an element of
$\widehat{\calq}^{(k+l-2)}$
with  a factor $\hbar^2$. This proves identities
  \eqref{genqucond1} and \eqref{genqucond2},
in particular, that
$Q|_{ (\calq^{(0)} \oplus\calq^{(1)})_\CC }:
\calq^{(0)}_\CC\oplus\calq^{(1)}_\CC\rightarrow \calqq^{(1)} $
is a Lie algebra isomorphism.  $\qquad\Box$

Theorem \ref{genqao} provides us with a way of constructing the
quantum algebra for Poisson algebras
$\widehat{\calq}=U(\gothe)\tenltimes \cif(E,\mathbb{C})$  with a semidirect
product Poisson structure
\eqref{genpb}. The next step is the study of their representation
theory, i.e. the classification of all irreducible Hilbert space
representations. For this, we must decide which representations we
want to consider and which meaning we want to give to the requirement of
irreducibility. These questions arise in a similar way in the standard
examples treated in textbooks on quantum mechanics, for the case of $T^*\RR^N$
 see
\cite{Straumann} and also the papers \cite{Gotay} for a more detailed
treatment.  Physical requirements usually result in restrictions on
the admissible representations, which are reflected in the concept of
integrability  and a specific interpretation of irreducibility.

The simplest case is that of a quantum algebra of observables which is
the universal enveloping algebra $U(\gothe)$ of a Lie algebra
$\gothe$. Via differentiation, representations of the corresponding
Lie group $E$ on a Hilbert space give rise to representations of
the universal enveloping algebra $U(\gothe)$ on a dense, invariant
subspace, the Garding space or space of $\cif$-vectors
\cite{Jorgensen}, \cite{BaRa}. However, in general not all
representations of the universal enveloping algebra arise that
way.
Following \cite{Jorgensen} we
call a representation $\Pi$ of $ U(\gothe)$
{\em integrable} if it is derived from a unitary Hilbert space representation  $\pi$ of $E$ according to the rule
\bea
\label{formalrule}
\Pi(\xi)\psi =
\frac{d}{dt}|_{t=0}\pi(\exp(-t \xi))\psi,  \qquad\quad \xi \in \gothg,
\eea
for all $\cif$-vectors $\psi$.
Because the elements of the Lie group $E$ often correspond to physically
meaningful transformations on the phase space, it is usually requested
that the Lie group $E$ be represented on the Hilbert space and only
integrable representations are considered. Similarly, irreducibility
is usually understood with respect to the representation of the Lie
group $E$, i.e. one classifies integrable representations of the
universal enveloping algebra $U(\gothe)$ for which the corresponding
representation of the Lie group $E$ is irreducible.

In our case, we want to impose analogous requirements for the
representations of the subalgebra $U(\gothe)\subset \widehat{\calq}$,
but we need to combine the associated representations of the group $E$
with representations of the function algebra $\cif(E,\mathbb{C})$. To do this,
 we recall that the product \eqref{genmult} in $\widehat{\calq}$ involves
the action \eqref{genpb} of $\gothe $ on $\cif(E,\mathbb{C})$ that is
derived from the  group action $.$ of $E$ on $\cif(E,\mathbb{C})$.
 Thus we want to
combine the group $E$ with the functions
$\cif(E,\mathbb{C})$ in such a way that the product of group elements and
functions  involves the action $g\in E$
which sends $F\in \cif(E,\mathbb{C})$ to $(u\mapsto F(g^{-1}.u))$. Tensoring group
elements with the function algebra $\cif(E,\mathbb{C})$
 only makes sense at the level
of the group algebra of $E$. If we realise the group algebra
of $E$ as (a certain class of)  functions on $E$ with multiplication
defined by convolution,  the combination with the function algebra
$\cif(E,\mathbb{C})$ can be
achieved in the framework of transformation group algebras, initiated
by Dixmier \cite{Dixmierart}
and continued by Glimm \cite{Glimm}. In the most general definition
of transformation group algebras
one starts with a topological group $E$
which is Hausdorff, locally compact and second countable and
acts   on a  space $X$. Here  we
only need to consider the case where $X=E$, and $E$ is a
finite-dimensional Lie group.
We summarise the key results
following  the paper
\cite{KM}, which gives a treatment that is closely related
to our situation.
\begin{definition}
\label{transalg}
Let $E$ be a unimodular   Lie group acting  continuously  on itself
via $.: E\times E\rightarrow E$. Then the space $C_0(E\times E,\mathbb{C})$
of continuous functions on $E\times E$ with compact support is a
transformation group algebra if it is equipped with
the multiplication and $*$-operation given by
\bea
\label{groupact}
F_1 \bullet  F_2 (v,u)&=& \int_E F_1(z,u) F_2(z^{-1}v,z^{-1}.u) \,\, dz \\
F^*(v,u)&=&\overline{F(v^{-1},v^{-1}.u)}.
\eea
With the norm $||F||_1=\int_E ||F(z,\cdot)||_\infty \,\,dz$
we have the inequality $||F_1 \bullet  F_2||_1\leq ||F_1||_1 ||F_2||_1$.
\end{definition}

We now define irreducible integrable  representations
 of $ \widehat{\calq}$ to be those which,
in  a sense to be specified below, are
derived from  irreducible  unitary and  bounded
representation of $C_0(E\times E,\mathbb{C})$.
The task of classifying integrable and irreducible representations
of  $ \widehat{\calq}$ then reduces, by definition, to the task
of classifying irreducible unitary and bounded representations of
the transformation algebra $C_0(E\times E,\mathbb{C})$.

Following the study  in \cite{KM},
we make the technical but important assumption
that  the orbit
space of the $E$-action on itself is $T_0$  in the quotient topology
(A topological space is $T_0$ if for any two distinct points at
least one of the points has a neighbourhood to which the other does
not belong). Also,
we assume  for simplicity the existence of an invariant measure $dm$
on the orbits of the group action of $E$ on itself. Then, the bounded
irreducible representations of a transformation group algebra are
characterised by the following theorem \cite{KM}.
\begin{theorem}
\label{transreps}
Let $E$ be as above and assume that the
orbit space of the $E$-action on itself is $T_0$ in the quotient
topology and that there exists
 invariant measures $dm$ on each orbit.
 Then the  irreducible $||\cdot ||_1$-bounded unitary representations
of the transformation group algebra $C_0(E\times E,\mathbb{C})$
 are labelled  by  orbits $O_{\mu}=\{v.g_{\mu}\,|\,v\in
E\}$, $g_{\mu}\in E$, of the action of $E$ on itself and
irreducible unitary representations $\Pi_{s}$ of the associated stabilisers
$N_\mu= \{n\in E\,|\,n.g_{\mu}=g_\mu\}$
on Hilbert spaces $V_s$.
The  representation spaces $V_{\mu s}$ are, up to equivalence,
given by the following construction.
Let  $L^2_{\mu s}$ be the space
\bea
L^2_{\mu s} := \{ \psi:E\rightarrow V_{s} \;|\; \psi(vn) =
\Pi_s (n^{-1})
\psi(v),\;\;\forall n\in N_\mu,\; \;\forall v \in E,\nonumber\\
\mbox{and} \quad
\|\psi\|^2:= \int_{E/N_\mu} \|\psi(z)\|_{V_s}^2\,dm(zN_\mu)<\infty\}
\eea
with the positive semi-definite inner product
\bea
\langle \psi_1,\psi_2\rangle:=
\int_{E/N_\mu}\langle \psi_1(z),\psi_2(z)\rangle_{V_s}\,d m (zN_\mu).
\eea
Then one obtains a Hilbert space by taking the quotient space with respect
to the subspace of functions with norm zero:
\bea
\label{genhilbert}
V_{\mu s}  = L^2_{\mu s} \: /\: \{\psi \in
L^2_{\mu s} \:|\: \|\psi\| = 0\},
\eea
on which elements $ F\in C_0(E\times E,\mathbb{C}) $   act via
\bea
\left(\pi_{\mu s}( F)\psi\right)\,(v)=\int_E F(z,v.g_\mu)\psi(z^\inv v)\,\,dz.
\eea
\end{theorem}

It remains to show that there is a dense
invariant
 subspace of the Hilbert space
$V_{\mu s}$ that carries a representation of the quantum algebra  $ \widehat{\calq}$  and to specify how this representation is
derived from that of $C_0(E\times E,\mathbb{C})$. For this purpose,
note that the group $E$ acts unitarily (and reducibly) on $V_{\mu s}$ via
\bea
\label{grouprep}
\left(\pi(g)\psi\right) (v) = \psi(g^{-1}v).
\eea
Following \cite{BaRa}, we define $\cif$-vectors in a representation
of $E$ to be those for which the map $\psi \mapsto \pi(g)\psi$ is
infinitely often differentiable. The $\cif$-vectors in $V_{\mu s}$
 viewed  as a representation
of $E$ are precisely those $\psi \in V_{\mu s}$ which are
smooth  functions $E\rightarrow V_s$. On these vectors
the derived action of the
Lie algebra $\gothe$ is then obtained by differentiation as in
\eqref{formalrule}.
 In order to obtain a subspace on
which $\cif(E,\mathbb{C})$ acts we need to impose the additional restriction that the
 map $\|\psi\|_{V_s}^2:E/N_\mu\mapsto \CC$ has compact support.
We define
\bea
\label{smoothrep}
V^\infty_{\mu s}=\{\psi \in  \calc^\infty(E,V_{s})|
\;\psi(v n)=\Pi_{s}(n^\inv)\psi(v)\quad\forall
n\in N_\mu, v\in E \nonumber \\
\mbox{and} \quad
||\psi||_{V_s}^2 \in \calc^\infty_0(E/N_\mu)\}
\eea
and have

\begin{theorem}
\label{irreps}
The space $V^\infty_{\mu s}$ is a dense subspace of the
Hilbert space  $V_{\mu s}$ and carries the derived  representation of
the  quantum algebra $\calqq$
defined by
\begin{align}
\label{genrep}
&\Pi_{\mu s}(\xi\otimes 1)\psi (v)=-i\hbar\,
\frac{\text{d}}{\text{d}t}|_{t=0}
\psi(e^{-t\xi}\,v)
& &\Pi_{\mu  s }(1\otimes F)\psi(v)=
F(v.g_\mu)\cdot \psi(v)
\end{align}
for $\xi \in\gothe$, $F\in \cif(E,\mathbb{C})$.
\end{theorem}

{\bf Proof:}
The density of $V^\infty_{\mu s}$ in $V_{\mu s}$ follows
from the density of $\calc^\infty_0(M,\mathbb{C})$ in $L^2(M,\mathbb{C})$ for any
domain $M$ \cite{BaRa}.
To see  that the action  \eqref{genrep}
 of $\calqq$ on  $V^\infty_{\mu s}$ is well-defined and
leaves $V^\infty_{\mu s}$  invariant note that if $||\psi||_{V_s}^2$
has compact support so does $|F(\cdot.g_\mu)|^2 ||\psi||_{V_s}^2$.
Checking that \eqref{genrep} defines a representation is an
easy algebraic exercise. $\hfill\Box$

\subsection{Quantisation of the decoupled flower algebra}
\label{decoupquantsect}
After constructing the quantum algebra and its representations
for Poisson algebras of type \eqref{poissalgdef} with brackets
 \eqref{genpb}
we can now apply these results to the cotangent bundle symplectic
structure $T^*G$, the dual Poisson structure $H^*$ and, finally,
the decoupled flower algebra $\gothf$.
For the case of the cotangent bundle $T^*G$ with the Poisson
brackets given by \eqref{hcotan}, the corresponding group action is
simply the inverse left multiplication or, alternatively, the right
multiplication with $G$. There is only a single orbit,
the group $G$, and its stabiliser group  is trivial. We obtain
the following quantum algebra
\begin{theorem} (Quantum algebra for $T^*G$)
\label{cotanalg}
\begin{enumerate}
\item The quantum algebra for the cotangent bundle Poisson structure
  $T^*G$ is the associative algebra
$U(\gothg)\tenltimes\cif(G,\mathbb{C})$
with the multiplication defined by
\begin{align}
\label{multhandle}
&(\xi\otimes F)\cdot(\eta\otimes K)=(\xi\cdot_U \eta)\otimes (FK)-i\hbar\;
\eta\otimes F( \xi^L K),
\end{align}
where $\xi,\eta\in\gothg$, $F,K\in\cif(G,\mathbb{C})$, $\xi^L$ is the right-invariant vector field associated to $\xi$ and
$\cdot_U$ denotes the multiplication in the universal enveloping
algebra $U(\gothg)$.

\item The corresponding transformation group algebra
 has a single irreducible representation  on the Hilbert  space
$V=L^2(G,\CC)$. Elements of   $U(\gothg)\tenltimes\cif(G,\CC)$ act on
the dense invariant subspace $\calc^\infty_0(G,\CC)$
 according to
\begin{align}
&\Pi(\xi\otimes 1)\psi (u)=-i\hbar\, \xi^L\psi(u) & &\Pi(1\otimes F)\psi(u)=
F(u)\cdot \psi(u)
\end{align}
for $u\in G, \xi\in\gothg$, $F\in\cif(G,\CC)$.
\end{enumerate}
\end{theorem}
Of course, we could just as well have $G$ let act on itself via the right multiplication and simply exchanged left and right in Theorem \ref{cotanalg}. Combining one copy of $T^*G$, where $G$ acts by right multiplication, and one where it acts by inverse left multiplication, we  obtain the quantum algebra
associated to the Poisson structure \eqref{hcotan}
of each handle:
\begin{definition} (Handle algebra $\widehat{\mathcal{H}}$)
\label{handledef}
\begin{enumerate}
\item The handle algebra is the associative algebra $
\widehat{\mathcal{H}}=U\left(\gothg\oplus\gothg\right)\tenltimes\cif(G\times G,\CC)$
with generators $k_{a}^1,k_a^2\in\gothg$
for the two copies of $\gothg$ and multiplication defined by
\begin{align}
\label{multhandle2}
&(\xi\otimes F)\cdot(\eta\otimes K)=(\xi\cdot_U \eta)\otimes (FK)+i\hbar\;
\eta\otimes F\{\xi\otimes 1,1\otimes K\},
\end{align} where  $\xi,\eta\in \gothg\oplus\gothg, F,K\in\cif(G\times G,\CC)$, $\cdot_U$ is the multiplication in $U(\gothg\oplus\gothg)$ and the bracket
$\{\,,\,\}$ is given by \eqref{hcotan}.
\item The corresponding transformation group algebra
 has a single irreducible representation
 on the space
  $L^2(G\times G,\CC)$ and elements of $\widehat{\mathcal{H}}$
act on the dense subspace $\calc^\infty_0(G\times G,\CC)$   via
\begin{align}
&\Pi_{\widehat{\mathcal  {H}}}( (k_a^1,k_b^2)\otimes 1)\psi
  (w_1,w_2)= -i\hbar\, \frac{\text{d}}{\text{d}t}|_{t=0}
  \psi(w_1e^{tJ_a},e^{-tJ_b}w_2),\\
&\Pi_{\widehat{\mathcal  {H}}}( 1\otimes F)\psi
  (w_1,w_2)=F(w_1,w_2)\cdot\psi(w_1,w_2).\nonumber
\end{align}
\end{enumerate}
\end{definition}
The case of the dual Poisson structure $H^*$ is slightly more
 complicated.  Here, the group action associated to the Poisson
 bracket is conjugation with $G$. Consequently, its orbits are
 $G$-conjugacy classes, and its irreducible representations are
 labelled by $G$-conjugacy classes and unitary irreducible representations 
of the associated  stabilisers.
\begin{theorem}  (Puncture algebra $\widehat{\mathcal{P}}$)
\label{punctth}
\begin{enumerate}
\item The quantum algebra $\widehat{\mathcal{P}}$ associated to the Poisson structure \eqref{partbr} on
  the dual Poisson-Lie group $H^*$, in the following referred to as
  puncture algebra,  is the associative algebra
$\widehat{\mathcal{P}}=U(\gothg)\tenltimes\cif(G,\CC)$
with multiplication defined by
\begin{align}
\label{multpunct}
&(\xi\otimes F)\cdot(\eta\otimes K)=(\xi\cdot_U \eta)\otimes (FK)-i \hbar\;
\eta\otimes F(\xi^L+\xi^R) K,
\end{align}
where $\xi,\eta\in\gothg$, $F,K\in\cif(G,\CC)$, $\xi^L$, $\xi^R$ are the right- and
left-invariant vector fields associated to $\xi$ and
$\cdot_U$ denotes the multiplication in the universal enveloping
algebra $U(\gothg)$.
\item The corresponding transformation group algebra is called the
quantum double of $G$ and denoted $D(G)$.
Under the technical assumptions of Theorem \ref{transreps}
 its irreducible representations, classified in \cite{KM},
 are labelled by the
  $G$-conjugacy classes $\calc_\mu=\{v\cdot g_\mu\cdot v^\inv\,| \, v\in G\}$
  and irreducible unitary
representations $\Pi_s$  of associated  stabilisers
$N_\mu=\{ n\in G| n\cdot g_\mu \cdot n^\inv=g_\mu\}$ on Hilbert spaces
$V_{s}$.
The representation spaces are
\bea
\label{doublerep}
V_{\mu s}= \{ \psi:G\rightarrow V_{s} \;|\; \psi(vn) =
\Pi_s (n^{-1})
\psi(v),\;\;\forall n\in N_\mu,\; \;\forall v \in G,\nonumber\\
\mbox{and} \quad
\|\psi\|^2:= \int_{G/N_\mu} \|\psi(z)\|_{V_s}^2\,dm(zN_\mu)<\infty\}/\sim,
\eea
where $\sim$ denotes division by zero-norm states and $dm$ is an
invariant
measure on $G/N_\mu$.
The algebra $\widehat{\mathcal{P}}$ acts on the dense subspace
$V^\infty_{\mu s}$ via
\bea
\Pi_{\mu s}(\xi\otimes 1)\psi (v)=-i\hbar \xi^L\psi(v)\qquad\Pi_{\mu s}
(1\otimes F)\psi (v)=F(v g_\mu v^\inv)\cdot\psi(v)
\eea
for $\xi\in\gothg$, $F\in\cif(G,\CC)$.
\end{enumerate}
\end{theorem}
After quantising the Poisson algebras \eqref{partbr} and \eqref{hcotan} associated to each puncture and handle, we can now combine these building blocks to construct the quantum algebra for the decoupled
 flower algebra and its irreducible representations. By inverting  transformation \eqref{hdecoup}, we obtain the quantum algebra $\hat{\gothf}$ associated to the Poisson algebra \eqref{partbr},\eqref{hbr}.
Note that  this Poisson algebra is again of type \eqref{poissalgdef}
with Poisson brackets \eqref{genpb}, where now
 $E=G^{n+2g}$ and the action of $G^{n+2g}$ on itself given by combining $n$ copies of the group action associated to the puncture algebra $\widehat{\mathcal{P}}$ and $g$ copies of the group action for the handle algebra $\widehat{\mathcal{H}}$:
\begin{align}
\label{floweract}
&(h_\me,\ldots,h_\mf,h_\aee,h_\bee,\ldots,h_\af,h_\bff)\,.\,
(u_\me,\ldots,u_\mf,u_\aee,u_\bee,\ldots,u_\af,u_\bff)\nonumber\\
&=(h_\me u_\me h_\me^\inv,\ldots,h_\mf u_\mf h_\mf^\inv,u_\aee h_\aee,u_\aee h_\aee u_\aee^\inv u_\bee h_\bee,\ldots,u_\af h_\af,u_\af h_\af u_\af^\inv u_\bff h_\bff).\nonumber
\end{align}

\begin{theorem}(Quantisation of the decoupled flower algebra)
\label{flowrep}
\begin{enumerate}
\item The quantum algebra for the decoupled flower algebra in Theorem \ref{decth}
  is the associative algebra
\bea
\label{flowernew}
\hat \gothf=U\left(\bigoplus_{k=1}^{n+2g} \gothg\right)
\tenltimes\cif(G^{n+2g},\CC),
\eea
with the multiplication defined by
\begin{align}
&(\xi\otimes F)\cdot(\eta\otimes K)=\xi\cdot_U \eta\otimes FK+i \hbar\;
  \eta\otimes F\{\xi\otimes 1,1\otimes K\},
\end{align}
where $\xi,\eta\in \bigoplus_{k=1}^{n+2g}\gothg, F,K\in\cif(G^{n+2g},\CC)$ and  $\cdot_U$ denotes the multiplication in
$U\left(\bigoplus_{k=1}^{n+2g}\gothg\right)$. The bracket $\{\,,\,\}$ is given by \eqref{partbr},\eqref{hbr} via the identification of the generators of the different copies of $\gothg$ with the elements  $j_a^{M'_1},\ldots,j_a^{M'_n}$, $j_a^{A'_1}-j_a^{B'_1},
j_a^{B'_1},\ldots,j_a^{A'_g}-j_a^{B'_g}, j_a^{B'_g}$ in Theorem
\ref{decth}. The algebra $\hat \gothf$ has a $*$-structure given by
$(j^{X'}_a)^*=j^{X'}_a$, $(1\otimes F)^*=1\otimes \bar{F}$.

\item  If the technical assumptions of Theorem \ref{transreps}
hold
the irreducible representations of the
 transformation
group algebra corresponding to \eqref{flowernew}  are labelled by $n$
  $G$-conjugacy classes $\calc_{\mu_1},\ldots,\calc_{\mu_n}$ and
 irreducible unitary
  representations $\Pi_{s_i}$
 of  stabilisers $N_{\mu_1},\ldots,N_{\mu_n}$  of
  chosen elements $g_{\mu_1},\ldots g_{\mu_n}$ in those conjugacy
  classes
on Hilbert  spaces $V_{s_i}$.
Let
\begin{align}
&L^2_{\mu_1s_1\ldots\mu_n s_n}=\big\{\psi: G^{n+2g} \rightarrow V_{s_1}
\otimes\ldots\otimes V_{
 s_n})\;|\;\psi( v_{M_1} h_1,\ldots,v_{M_n} h_n,  u_{A_1},\ldots, u_{B_g})\nonumber\\
&\quad=(\Pi_{s_1}(h_1^\inv)\otimes\ldots\otimes
  \Pi_{s_n}(h_n^\inv))\;\psi(v_{M_1},\ldots,
  v_{M_n},u_{A_1},\ldots,u_{B_g})\; \forall h_i\in    N _{\mu_i},\;
  ||\psi||^2 <\infty\big\},\nonumber
\end{align}
with norm $||\cdot ||$ derived from the inner product
\begin{align}
\langle \psi,\phi\rangle=\int_{G/N_{\mu_1}\times\ldots\times G/N_{\mu_n}\times G^{2g} } &\left(\,\psi\,,\, \phi\,\right)
(v_{M_1},\ldots,v_{M_n},u_{A_1},\ldots u_{B_g})\\
&\qquad\qquad dm_1(v_{M_1}\cdot N_1)\cdots
dm_n({v_{M_n}\cdot N_n})\cdot du_{A_1}\cdots du_{B_g},\nonumber
\end{align}
where  $(\,,\,)$  is the canonical inner product on the tensor
product of Hilbert spaces $V_{s_1}\otimes \ldots\otimes V_{s_n}$.
The representation spaces are
\bea
\label{repspace}
V_\repind=L^2_{\mu_1s_1\ldots\mu_n s_n}/\sim
\eea
where $\sim$ denotes division by zero-norm states.
Elements of $\hat \gothf$ act on the dense subspace
$V^\infty_{\mu_1s_1\ldots\mu_n s_n}$
according to
\begin{align}
\label{quantact}
&\Pi_\repind(1\otimes F)\psi(v_{M_1},\ldots, v_{M_n},u_\aee,\ldots,u_\bff)\\
&\qquad\qquad=F(v_{M_1}g_{\mu_1}
v_{M_1}^\inv, \ldots,v_{M_n}g_{\mu_n}
v_{M_n}^\inv, u_{A_1},\ldots, u_{B_g} )\Psi(v_{M_1},\ldots,v_{M_n},u_\aee,\ldots,u_\bff)\nonumber\\
&\qquad\qquad=((F\circ\beta)\cdot\psi)(v_{M_1},\ldots, v_{M_n},u_\aee,\ldots,u_\bff)\nonumber\\
&\Pi_\repind(j^\mip_a\otimes 1)\psi(v_{M_1},\ldots,
v_{M_n},u_\aee,\ldots,u_\bff)\nonumber\\
&\qquad\qquad=-i\hbar\, \frac{d}{dt}|_{t=0}\psi(v_{M_1},\ldots,
e^{-tJ^a}v_\mi,\ldots, v_{M_n},u_\aee,\ldots,u_\bff)\nonumber\\
&\qquad\qquad=-i\hbar\, J_a^{L_\mi}\,\psi(v_{M_1},\ldots,v_{M_n},u_\aee,\ldots,u_\bff)\nonumber\\
&\Pi_\repind((j^\aip_a-j_a^\bip)\otimes 1)\psi(v_{M_1},\ldots,
v_{M_n},u_\aee,\ldots,u_\bff)\nonumber\\
&\qquad\qquad=-i\hbar\, \frac{d}{dt}|_{t=0}\psi(v_{M_1},\ldots,u_\ai e^{tJ^a}, u_\ai
e^{tJ^a}u_\ai^\inv u_\bi,\ldots, u_{B_g})\nonumber\\
&\qquad\qquad=-i\hbar\, (J_a^{R_\ai}+ \Ad^*(u_\ai^\inv u_\bi)_{a}^{\;\;b}J_b^{R_\bi})\,\psi(v_{M_1},\ldots,v_{M_n},u_\aee,\ldots,u_\bff)\nonumber\\
&\Pi_\repind(j_a^\bip\otimes 1)\psi(v_{M_1},\ldots,
v_{M_n},u_\aee,\ldots,u_\bff)\nonumber\\
&\qquad\quad=-i\hbar\, \frac{d}{dt}|_{t=0}\psi(v_{M_1},\ldots,u_\ai, u_\bi e^{tJ^a},\ldots, u_{B_g})\nonumber\\
&\quad\qquad=-i\hbar\, J_a^{R_\bi}\,\psi(v_{M_1},\ldots,v_{M_n},u_\aee,\ldots,u_\bff),\nonumber
\end{align}
 with $\beta: G^{n+2g}\rightarrow G^{n+2g}$ given by
\bea
\label{betadef}
\beta(v_{M_1},\ldots, v_{M_n},u_\aee,\ldots,u_\bff):=(v_{M_1}g_{\mu_1}
v_{M_1}^\inv, \ldots,v_{M_n}g_{\mu_n}
v_{M_n}^\inv, u_{A_1},\ldots, u_{B_g}).
\eea
\item
  With the induced  inner product on the subspace $V^\infty_\repind$,
the representations \eqref{quantact} are
 $*$-representations with respect to the $*$-structure given above and
the operators  $\Pi_\repind(j^{X'}_a\oo 1)$, $\Pi_\repind(1\otimes F)$ are
Hermitian.
\end{enumerate}
\end{theorem}

\subsection{The quantum decoupling transformation}

We can now use the quantisation of the decoupled Poisson structure in Theorem \ref{flowrep} to obtain the quantum version of the original, undecoupled Poisson brackets \eqref{poiss} of the flower algebra. The idea is to
 define a quantum counterpart of the inverse decoupling transformation
 $K^\inv$ given by \eqref{invdecoup}. In trying to implement this
 idea
one encounters ordering ambiguities in
 the part of $K^\inv$ that involves the generators $j_a^\ai, j_a^\bi$
 associated to the handles, such that the quantum versions of the
 brackets  \eqref{poiss} associated to different choices of ordering
 would differ by quantum corrections. However, a closer look at the
 structure of the quantum algebra $\hat \gothf$ provides us with a
 canonical definition of the inverse quantum decoupling
 transformation. Note that the classical decoupling transformation and
 its inverse are {\em linear} in the generators $j_a^{X}$ and
 $j_a^{X'}$. We can therefore interpret them  as
bijective maps on the vector space
$\gothf^{(0)}\oplus\gothf^{(1)}$ and use the vector space isomorphism
 $Q$ in Theorem \ref{genqao} to define its quantum
counterpart.

\begin{theorem}(Quantum decoupling transformation)

\label{qudecouple}
Define the quantum decoupling transformation as
\bea
\label{qdecoup}
\hat K:= Q|_{01}\circ K\circ Q|_{01}^\inv:\; \hat\gothf^{(1)}\rightarrow\hat\gothf^{(1)},
\eea
where  $Q|_{01}$ denotes the map
$Q|_{(\gothf^{(0)}\oplus\gothf^{(1)})_\CC}:\gothf^{(0)}_\CC\bigoplus\gothf^{(1)}_\CC
\rightarrow \hat{\gothf}^{(1)}$ to
simplify notation, and
transform the generators  of $\hat \gothf^{(1)}$ with its inverse
$\hat K^\inv=Q|_{01}\circ K^\inv\circ Q|_{01}^\inv$. Then  the commutators of the transformed generators are given by applying the map $i\hbar \;Q|_{01}$ to the right-hand side of equations \eqref{poiss}.
\end{theorem}
Note that, although this construction looks quite formal, it
 amounts to the choice of an ordering in
\eqref{transfj},\eqref{invdecoup}, namely ordering all the generators
$j_a^X$, $j_a^{X'}$ that occur in these expressions to the right. The quantisation
of the Poisson brackets \eqref{poiss} obtained this way is canonical
in the following sense. Although the right-hand sides of \eqref{poiss}
contain products of generators $j_a^Y\in\bigoplus_{k=1}^{n+2g}\gothg$ with functions  $\Ad^*(u_X)_b^{\;\,c}\in\cif(G^{n+2g})$, these products do not give rise to ordering ambiguities. This is due to the fact that $X<Y$ in all of them, for which the last three brackets in \eqref{poiss} imply that the factors commute.

\section{Quantum symmetries and the quantum double $D(G)$}
\label{quantsymm}

\subsection{Quantum action of $\defgr$}
\label{quantsymm1}

With the definition of the quantum flower algebra $\hat \gothf$ and
its irreducible representations in Theorems \ref{flowrep} and
\ref{qudecouple}, we can now determine how  the group
$G\ltimes\cif(G)$ acts on this algebra.
The crucial observation for constructing the quantum action of this symmetry group is the fact that it induces  {\em linear} transformations of the classical generators $j_a^X$, as explained in Theorem \ref{defconjact}. This allows us to define their action on the generators of the quantum algebra by means of the map $Q$ in Theorem \ref{genqao}:

\begin{theorem}(Action of $\defgr$ on $\hat\gothf$)
\label{symth}

For an element $(h,f)\in G\ltimes\cif(G)$, define its
 action $\widehat{(h,f)} $ on the subalgebra $\hat
\gothf^{(1)}$ by
\bea
\label{quconj}
\widehat{(h,f)}=Q|_{01}\circ (h,f)\circ Q|_{01}^\inv.
\eea
This map  is  a Lie algebra automorphism of
$\hat \gothf^{(1)}$ and can be extended canonically to
an algebra isomorphism
 $\widehat{(h,f)}:\hat \gothf\rightarrow\hat\gothf$
of the quantised flower algebra.
\end{theorem}

{\bf Proof:} Let $\tau$ be a Poisson isomorphism of the classical
flower algebra that restricts to a Lie algebra automorphism of the
linear subalgebra $\gothf^{(0)}_\CC\oplus\gothf^{(1)}_\CC$ and therefore defines a Lie algebra automorphism $\hat\tau:\hat \gothf^{(1)}\rightarrow\hat\gothf^{(1)}$ by $\hat\tau=Q|_{01}\circ\tau\circ Q|_{01}^\inv$. As $Q|_{01}: \gothf^{(0)}_\CC\oplus\gothf^{(1)}_\CC \rightarrow
\hat\gothf^{(1)}$ is a Lie algebra isomorphism and $\tau$ a Poisson
automorphism of $\gothf$, we have
\begin{align}
[\hat\tau(\theta),\hat\tau(\chi)]&=i\hbar\; Q|_{01}(\{\tau(Q|_{01}^\inv(\theta)),\tau(Q|_{01}^\inv(\chi))\})\\
&=i\hbar\;
Q|_{01}\circ\tau(\{Q|_{01}^\inv(\theta),Q|_{01}^\inv(\chi)\})\nonumber\\
&=Q|_{01}\circ\tau\circ Q|_{01}^\inv([\theta,\chi])=\hat\tau([\theta,\chi])\qquad\forall \theta,\chi\in\hat\gothf^{(1)}\nonumber.
\end{align}
Via the choice of an ordered basis of
$U\left(\bigoplus_{k=1}^{n+2g}\gothg\right)$, for example the ordered polynomials in the elements of the ordered basis $j^{M'_1}_1<\ldots<j^{M'_1}_{\text{dim}\,G}<j^{M'_2}_1<\ldots<j^{M'_n}_{\text{dim}\,G}<j^{A'_1}_1<\ldots<j^{A'_1}_{\text{dim}\,G}<j^{B'_1}_1<\ldots<j^{B'_g}_{\text{dim}\,G}$  of the Lie algebra $\bigoplus_{k=1}^{n+2g}\gothg$, and setting
\begin{align}
&\hat\tau((j^{X'_1}_{a_1}\cdots j^{X'_m}_{a_m})\otimes 1):=\hat\tau(j^{X'_1}_{a_1}\otimes 1)\cdots\hat\tau(j^{X'_m}_{a_m}\otimes 1)\\
&\hat\tau((1\otimes F)\cdot(\theta\otimes 1)):=\hat\tau(1\otimes F)\cdot\hat\tau(\theta\otimes 1)\nonumber
\end{align}
for elements $\theta$ of this ordered basis, $\hat\tau$ can be extended to a
vector space isomorphism on $\hat\gothf$. Because $\hat\tau$ has been
extended multiplicatively to the ordered basis and $[\hat\tau(\xi\otimes
F),\hat\tau(\eta\otimes K)]=\hat\tau([\xi\otimes F,\eta\otimes K])$ for $\xi,\eta\in\bigoplus_{k=1}^{n+2g}\gothg$ and $F,K\in\cif(G^{n+2g},\CC)$, $\hat\tau:\hat\gothf\rightarrow\hat\gothf$ is also an algebra isomorphism.$\hfill\Box$

The action of the group $G\ltimes\cif(G)$ as an algebra isomorphism of the
quantised flower algebra raises the question if it
can be implemented unitarily  in  the representation spaces \eqref{repspace}.
 The following theorems show that this is indeed possible and give
explicit formulae for the action of $G\ltimes\cif(G)$
in  the representation spaces \eqref{repspace}.

\begin{theorem} (Representations of $\defgr$)
\label{defgrrep}

Let $\Pi_\repind$ be a
representation of the flower algebra as defined in Theorem \ref{flowrep}. Let the maps $\beta, \widetilde{Ad}_h^{n+2g}: G^{n+2g}\rightarrow G^{n+2g}$ be given by \eqref{betadef} and
\begin{align}
\label{addef}
&\widetilde{\Ad}_h^{n+2g}(v_{M_1},\ldots,v_{M_n},u_\aee,\ldots,u_\bff)=(h v_{M_1},\ldots,h v_{M_n},h u_{A_1}h^\inv,\ldots, hu_{B_g}h^\inv).
\end{align}
Then $\Gamma:\defgr\rightarrow\text{End}(V_\repind)$
\begin{align}
\label{defrep}
&\Gamma(h,f)\Psi(v_{M_1},\ldots,v_{M_n},u_\aee,\ldots,u_\bff)\\
&\qquad\qquad =\left(e^{\frac{i}{\hbar}f\circ\Phi_\infty\circ\beta}\cdot (\Psi\circ\widetilde{\Ad}_{h^\inv}^{n+2g})\right)(v_{M_1},\ldots,v_{M_n},u_\aee,\ldots,u_\bff)\nonumber\\
&\qquad\qquad=e^{\frac{i}{\hbar}f(u_{tot})}\, \Psi(h^\inv v_{M_1},\ldots, h^\inv v_{M_n}, h^\inv u_{A_1}h,\ldots, h^\inv u_{B_g}h)\nonumber
\end{align}
with $u_{tot}$ given by by \eqref{utot}, defines a representation of the group $\defgr$ on the representation space $V_\repind$ that satisfies
\bea
\label{repid}
\Pi_\repind(\widehat{(h,f}) Z)=\Gamma(h,f)\circ\Pi_\repind(Z)\circ\Gamma^\inv(h,f) \qquad\forall Z\in\hat\gothf
\eea
on the dense invariant subspace $V^\infty_\repind$.
If the conditions in theorem \ref{flowrep} are fulfilled,
this representation is unitary.
\end{theorem}

{\bf Proof:} To simplify notation, we write $V$ for $V_\repind$ and $\Pi$ for $\Pi_\repind$. The identity  $f\circ\Phi_\infty\circ\beta\circ\widetilde{Ad}_{h^\inv}^{n+2g}=f\circ\Ad_{h^\inv}\circ\Phi_\infty\circ\beta$ implies that
 \eqref{defrep} defines a representation of $\defgr$ on $V$. Since the
 group elements corresponding to the punctures are left-multiplied by
 elements of $G$ and the elements corresponding to the handles
 conjugated in \ref{addef}, the conditions on the measures in Theorem \ref{flowrep} guarantee unitarity for the representations of elements $(h,0)\in\defgr$. For elements $(1,f)\in\defgr$, which act via multiplication by a phase, this is trivial.

 To prove identity \ref{repid}, we calculate
\begin{align}
&\Pi\big(\widehat{(h,f)}(1\otimes K)\big)\circ \Gamma(h,f)\Psi= \Pi\big(1\otimes (K\circ\Ad^{n+2g}_{h^\inv})\big)\,\big(\, e^{\frac{i}{\hbar}f\circ\Phi_\infty\circ\beta}\cdot (\Psi\circ\widetilde{\Ad}_{h^\inv}^{n+2g})\,\big)\\
&=(K\circ\Ad^{n+2g}_{h^\inv}\circ\beta)\cdot e^{ \frac{i}{\hbar}f\circ\Phi_\infty\circ\beta}\cdot (\Psi\circ\widetilde{\Ad}_{h^\inv}^{n+2g})=\Gamma(h,f)\circ\Pi(1\otimes K)\Psi\qquad\forall\Psi\in V\nonumber
\end{align}
and, with the Poisson bracket \eqref{partbr},\eqref{hbr} for the generators $j_a^{X'}$ of the decoupled flower algebra
\begin{align}
&\Pi(\widehat{(h,f)}j_a^{X'})\circ \Gamma(h,f)\Psi= \Pi\big( (1\otimes\Ad^*(h)_a^{\;\;b})\cdot j_b^{X'}+1\otimes \Ad^*(h)_a^{\;\;b}\{j_b^{X'},f\circ\Phi_\infty\}\big)(\Gamma(h,f)\Psi)\nonumber\\
&=\Ad^*(h)_a^{\;\;b}\,(\,\{j_b^{X'},f\circ\Phi_\infty\}\circ\beta\,)\cdot
e^{\frac{i}{\hbar}f\circ\Phi_\infty\circ\beta}\cdot
(\Psi\circ\widetilde{\Ad}_{h^\inv}^{n+2g})\\
&\quad+\Ad^*(h)_a^{\;\,b}\, e^{\frac{i}{\hbar}f\circ\Phi_\infty\circ\beta}\cdot\big(\Pi(j_b^{X'})(\Psi\circ\widetilde{\Ad}_{h^\inv}^{n+2g})\big)\nonumber\\
&\quad+\tfrac{i}{\hbar} \Ad^*(h)_a^{\;\,b}\,e^{\frac{i}{\hbar}f\circ\Phi_\infty\circ\beta}\cdot(\Psi\circ\widetilde{\Ad}_{h^\inv}^{n+2g})\cdot\big(\Pi(j_b^{X'})(f\circ\Phi_\infty\circ\beta)\big)\nonumber\\
&= e^{\frac{i}{\hbar}f\circ\Phi_\infty\circ\beta}\cdot(\Pi(j_a^{X'})\Psi)\circ\widetilde{\Ad}_{h^\inv}^{n+2g}=\Gamma(h,f)\circ\Pi(j_a^{X})\Psi\qquad\forall \Psi\in V\nonumber,
\end{align}
where we used the identities
\begin{align}
&\Ad^*(h)_a^{\;\,b}\,\Pi(j_b^{X'})(\Psi\circ\widetilde{\Ad}_{h^\inv}^{n+2g})=\big(\Pi(j_b^{X'})\Psi\big)\circ\widetilde{\Ad}_{h^\inv}^{n+2g}\\
&\Pi(j_b^{X'})(f\circ\Phi_\infty\circ\beta)=i\hbar\,\{j_b^{X'}\otimes 1,f\circ\Phi_\infty\}\circ\beta\nonumber.
\end{align}
Therefore, we have $\Pi(\widehat{(h,f))} \theta)=\Gamma(h,f)\circ\Pi(\theta)\circ \Gamma^\inv(h,f)$ for all $\theta\in \gothf$.$\hfill\Box$

\subsection{The relation to the quantum double $D(G)$}

\label{qudouble}

In the combinatorial approach to quantising Chern-Simons theory
\cite{AGSI, AGSII, BR, BNR} quantum groups or, more precisely,
ribbon-Hopf-*-algebras and associated
structures play a central role. Since our approach to quantising
Chern-Simons theory with gauge group $\prgr$ begins with
a description of the classical phase space which is
analogous to that used in \cite{AGSI, AGSII} it is perhaps surprising the
we arrived at a quantisation without making explicit use of quantum group
theory so far.

In this section we discuss which role quantum groups, more precisely,
the quantum double $D(G)$ of the Lie group $G$
play in our formalism. We already encountered $D(G)$ as the
transformation group algebra associated with the puncture algebra
in Theorem \ref{punctth}. Here we will exhibit its ribbon-Hopf
properties.
We show that the quantum double $D(G)$ also acts on the representation
spaces of the handle algebra $\widehat{\mathcal{H}}$ given
in Def.~\ref{handledef}. Using this  action
 we obtain a representation of the quantum double on the representation spaces of the quantised flower algebra defined in Theorem \ref{flowrep}. By comparison with the quantum symmetries discussed in Sect.~\ref{quantsymm1}, it then becomes apparent that the quantum double can be viewed as a generalisation of the symmetry group $G\ltimes\cif(G)$.

The quantum double of a Lie group $G$  has been studied  in
various publications. We adopt  the conventions used in \cite{KM}
where the quantum double is identified with continuous
functions on $G\times G$,
except that we exchange the roles
played by the two copies of $G$ in order to match our
conventions for the semidirect product group $H$.
Thus we identify $D(G) =C_0(G\times G,\CC)$ as a vector space.
In order to exhibit the structure of $D(G)$ as a ribbon-Hopf-*-algebra,
we need to include Dirac delta functions which are not strictly in
$C_0(G\times G,\CC)$. One can avoid this problem by modifying the definition
of $D(G)$ as explained in \cite{schroers} or by simply adjoining
the singular elements. In practice the latter approach is more
convenient. Later we shall see that it  is precisely the singular elements
$\delta_g\otimes f\,\,(v,u)=\delta_g(v) f(u)$ which have a  conceptually simple
interpretation.

Thus we define   multiplication $\bullet$, identity 1,
co-multiplication $\Delta$,
co-unit $\epsilon$, antipode $S$ and involution ${}^*$ via
\bea
\label{algebra}
(F_1\bullet F_2)(v,u)&:=&\int_G F_1(z,u)\,F_2(z^{-1}v,z^{-1}uz)\,dz,
 \\
1(v,u)&:=&\delta_e(v), \\
(\Delta F)(v_1,u_1;v_2,u_2)&:=&F(v_1,u_1u_2)\,\delta_{v_1}(v_2). \\
\epsilon(F)&:=&\int_G F(v,e)\,dv, \\
(S F)(v,u)&:=&F(v^{-1},v^{-1}u^{-1}v), \\
F^*(v,u)&:=&\overline{F(v^{-1},v^{-1}uv)},
\eea
so that we have for the  singular   elements
\begin{align}
\label{singdoublemult}
&(\delta_{g_1}\otimes  f_1)\bullet (\delta_{g_2}\otimes f_2)
= \delta_{g_1g_2}\otimes ( f_1\cdot f_2\circ\Ad_{g_1^\inv})\\
&\Delta(\delta_g\otimes f)(v_1,u_1;v_2,u_2)=\delta_g(v_1)\delta_{g}(v_2) f(u_1u_2)\\
&\epsilon(\delta_g\otimes f)=f(e)\\
&S(\delta_g\otimes f)=\delta_{g^\inv}\otimes (f\circ\Ad_{g^\inv}\circ(\,)^ \inv)\\
\label{starop}
&(\delta_g\otimes f)^*=\delta_{g^\inv}\otimes(\overline{f}\circ\Ad_{g^\inv}).
\end{align}
The universal
$R$-matrix of  $D(G)$ is
\bea
R(v_1,u_1;v_2,u_2) = \delta_e(v_1)\delta_e(u_1v_2^{-1})
\eea
and the central ribbon element $c$
\bea
\label{randc}
c(v,u) =  \delta_v(u).
\eea
It  satisfies the ribbon relation
\bea
\label{ribbon}
\Delta c= (R_{21}\bullet R) \bullet \bigl(  c\otimes c \bigr)
\eea
with $R_{21}(v_1,u_1;v_2,u_2):=R(v_2,u_2;v_1,u_1)$.

The irreducible representations of $D(G)$  are given in Theorem
 \ref{punctth}.
 With the notation introduced
there, the  singular elements have the simple action
\bea
\Pi_{\mu s}(\delta_g\otimes f)\psi (v)= f(vg_\mu v^{-1}) \psi(g^{-1}v).
\eea
There is a further representation of $D(G)$ which will be relevant in the following, but which is not irreducible. It is obtained by letting
$D(G)$ act on itself via the adjoint action. Using Sweedler notation as defined in \eqref{sweedlermult}
the  adjoint action of $F\in D(G)$ on $\phi\in D(G)$ is
\bea
\label{adrep}
\ad(F) \phi (w_1,w_2)&=&\sum F_{(1)}\bullet\phi \bullet SF_{(2)}
\nonumber \\
&=& \int_G\,\,F(v,w_1w_2^{-1}w_1^{-1}w_2) \phi(v^{-1}w_1v,v^{-1}w_2v)
\,dv
\eea implying
\bea
\label{singdoublerep}
\ad(\delta_g\otimes k)
\phi (w_1,w_2)
= k(w_1w_2^{-1}w_1^{-1}w_2) \phi(g^{-1}w_1g,g^{-1}w_2 g).
\eea

Now note that the same action can be used to let $D(G)$
act on the irreducible representation of the handle algebra
$\widehat{\mathcal{H}}$. Combining the representations \eqref{doublerep}
and \eqref{adrep} and using the co-multiplication
of $D(G)$ repeatedly  we obtain an action of $D(G)$ on the
 representation spaces $V_\repind$ of the flower algebra:
\bea
\Pi(F)\psi=(\Pi_{\mu_1s_1}\otimes\cdots\otimes\Pi_{\mu_ns_n}\otimes\ad\otimes\cdots\otimes\ad)\circ(\Delta\otimes\underbrace{1\otimes\cdots\otimes 1}_{(n+g-2)\times})\circ\ldots\nonumber\\
\ldots\circ(\Delta\otimes 1\otimes 1)\circ(\Delta\otimes 1)(F)\psi,\label{symact}.
\eea
For the singular elements we find
\begin{align}
\label{symact2}
&\Pi(\delta_h\otimes k) \psi(v_{M_1},\ldots, v_{M_n}, w_{1,1},w_{2,1},\ldots
,w_{1,g},w_{2,g})\\
&\qquad\qquad
=(k\circ\Phi_\infty\circ\beta)
\cdot(\psi\circ
\widetilde{\Ad}_{h^\inv}^{n+2g})
(v_{M_1},\ldots, v_{M_n}, w_{1,1},w_{2,1},\ldots
,w_{1,g},w_{2,g})\nonumber\\
&\qquad\qquad= k(u_{tot}) \psi(h^\inv v_{M_1},\ldots, h^\inv v_{M_n}, h^{-1} w_{1,1} h, h^{-1} w_{2,1} h,
\ldots  ,h^{-1}w_{1,g}h ,h^{-1} w_{2,g}h),\nonumber
\end{align}
where  $\beta$, $\widetilde{\Ad}_{h^\inv}^{n+2g}$ are
given by \eqref{betadef}, \eqref{addef} and $\Phi_\infty$ and
$ u_{tot}$ by \eqref{utot} with the identification
$w_{1,i}=u_\ai$, $w_{2,i}=u_\bi^\inv u_\ai$ as in \eqref{hdecoup}.

Comparing this representation of the quantum double $D(G)$ on the
representation space $V_\repind$ with the quantum action
\eqref{defrep} of the group $G\ltimes\cif(G)$ in Theorem
\ref{defgrrep}, we see that they are identical if we identify
$h\leftrightarrow \delta_h$ and $k\leftrightarrow e^{
  \frac{i}{\hbar}f}$. Furthermore, with this identification, the
multiplication law \eqref{singdoublemult} of $D(G)$ agrees with the
multiplication \eqref{defmult} of the group $\defgr$ and the
$*$-operation \eqref{starop} maps each element of $\defgr$ to its
inverse. In other words, the map
\bea
(h,f)\mapsto \delta_h \otimes  e^{\frac{i}{\hbar}f}
\eea
is a group morphism from the symmetry group $\defgr$
into the semidirect product  $G\ltimes\cif(G, U(1))$ of
$G$ with smooth $U(1)$-valued functions on $G$,
realised as a group
of (singular) elements in   the quantum double $D(G)$.
In this sense the  quantum group $D(G)$
generalises the   symmetry group $\defgr$.
Moreover, the formula  \eqref{symact} shows  that the action
of this generalised symmetry on the Hilbert space $V_\repind$
is naturally expressed in terms
of the  co-multiplication
of $D(G)$.

This relation between the group $\defgr$ and the quantum double $D(G)$ fits in nicely with the role quantum groups play in the formalism of combinatorial quantisation of Chern-Simons gauge theories.
 In Sect.~\ref{flowalg}, we explained how the phase space of
 Chern-Simons gauge theory with gauge group $H$, the moduli space of
 flat $H$-connections, is related to the classical flower algebra. It
 is obtained from the space of holonomies by dividing out the residual
 gauge transformations that act on the holonomies by global
 conjugation with $H$, i.e. by imposing the constraint arising from
\eqref{pirel}.
 In the combinatorial quantisation scheme \cite{AGSI, AGSII, AS}, the representation spaces of the quantised moduli space are then constructed by imposing invariance under the action of a corresponding quantum group on the representation spaces of the quantum flower algebra.

In our formalism, the constraint $(u_{tot}, -\Ad(u_{tot})\bj)\approx
1$ arising from \eqref{pirel} appears  as the infinitesimal generator
of the classical action of the symmetry group $\defgr$ in Theorem
\ref{defconjact}. Its action on the flower algebra can be interpreted
as a generalised or deformed conjugation. Implementing this constraint
via the Dirac formalism \cite{AGSI, AGSII, AS} would then amount to selecting the states on the representation spaces $V_\repind$ of the quantum flower algebra that are invariant under the action of the group $\defgr$.

\section{Outlook and conclusions}

In this paper we quantised the flower algebra, which is a crucial
ingredient in the description of the phase space of
Chern-Simons gauge theory with semidirect product gauge groups
$H=\prgr$ on a punctured  surface.
 We showed how its Poisson structure
can be broken up into a set of Poisson commuting
building blocks and discussed the Poisson action of the group $\defgr$.
 This allowed us to construct the corresponding
quantum algebra  and its irreducible Hilbert space representations
by means of a rather straightforward quantisation procedure. After determining
the action of the group $\defgr$ on the quantum algebra, we
were then able to relate this group action to the quantum
double $D(G)$  of the Lie group $G$. This clarified how
this quantum group arises as a quantum symmetry.

It is interesting to compare our approach to
the formalism of combinatorial
quantisation of Chern-Simons gauge theories developed for
compact, semisimple gauge groups in \cite{AGSI, AGSII, AS} and its
extension to the case of the semisimple but non-compact group $SL(2,\mathbb{C})$
in \cite{BR, BNR}.
Both start from the classical flower algebra and in both
cases quantum groups play an important role. However,
it is not clear how
the  combinatorial quantisation scheme could be generalised
to groups of the form $H=\prgr$ in a mathematically rigorous
fashion.
For this reason, we did not use it
 as a guideline for quantisation but based our approach
on a detailed investigation of the structure of the classical
algebra.

We see explicitness and simplicity as an advantage of our approach. It
describes the classical flower algebra in terms of quantities that can
easily be related to the physical content of the underlying
Chern-Simons gauge theory. For instance, in the case of
(2+1)-dimensional gravity in its formulation as a Chern-Simons theory
with the three-dimensional Poincar\'e group as gauge group, our
parameters represent momenta and angular momenta of handles and
massive
particles with spin \cite{we}. All of the structural properties of the
flower algebra, the action of the symmetry group $\defgr$ as well as
the transformation that decouples the contributions of different
punctures and handles, can be expressed in terms of these quantities.
This allows us to perform concrete calculations and to gain insight
into their physical interpretation. Similarly, the corresponding
quantum algebra is given explicitly as a semidirect product of an
universal enveloping algebra and an abelian algebra of functions,
rather than implicitly by a set of generating matrix elements and
relations as in the combinatorial quantisation formalism.
This facilitates the investigation of its structure and the
study of its representation theory.

We did not  impose the constraint that the holonomies
around punctures lie in fixed $H$-conjugacy classes
either in the classical or in the quantised flower algebra,
mainly for technical reasons.
Instead, we
found that the irreducible representations of the quantised flower
algebra correspond to the
symplectic leaves of the classical flower algebra. Recall that the
latter  are of the form $\calc_{\mu_1 s_1}\times \ldots \times
\calc_{\mu_n s_n}\times H^{2g}$, where $\mu_i$ label $G$-conjugacy
classes and $s_i$ co-adjoint orbits of associated stabiliser groups.
The irreducible representations \eqref{repspace}
 of the quantised flower algebra are labelled by  $G$-conjugacy
classes and irreducible representations of associated stabiliser groups.
The correspondence between symplectic leaves and irreducible
representations is thus the familiar correspondence between
co-adjoint orbits and irreducible representations \cite{Kirillov}
which typically holds for quantised values of the parameters
labelling the co-adjoint orbits. Details depend on the group $G$
and seem worth investigating further. In particular,
one should be able to obtain the representation
spaces $V_{\mu s}$ of the puncture algebra directly by geometric
quantisation of the conjugacy classes $\calc_{\mu s}$. For
an approach to the quantisation of closely related spaces  $(T^*G)/N$
via $C^*$-algebras see \cite{Landsman}.

The quantisation of the flower algebra for Chern-Simons gauge theories
with gauge groups $H=\prgr$ constitutes an important step towards the
quantisation of the moduli space of flat $H$-connections. To obtain a
quantisation of the moduli space from this quantisation of the flower
algebra, one would have to implement the constraint arising from \eqref{pirel}
which acts as the infinitesimal generator of the action of the group
$\defgr$. Doing this via the Dirac quantisation procedure would amount
to determining the subspaces of the representation spaces of the
flower algebra that are invariant under the action of $\defgr$ or the
quantum double $D(G)$. This requires a Clebsch-Gordon analysis of
tensor product representations of the
quantum double $D(G)$. For compact groups $G$, a general framework for
doing this was developed in \cite{KBM}, but
explicit calculations of Clebsch-Gordon coefficients depend on the
particular choice of $G$. The case  $G=SU(2)$
was studied in \cite{BM}. For other groups such as the group
$G=\widetilde{SO(2,1)}$ occurring in (2+1)-dimensional gravity, this
remains an open question and possible subject of further investigations.

\section*{Acknowledgements}
CM acknowledges financial support by the Engineering and Physical
Sciences Research Council and a living stipend from the
Studienstiftung des Deutschen Volkes. BJS
thanks Tom Koornwinder, Klaas Landsman and Joost Slingerland
for helpful discussions
and acknowledges an Advanced
Research Fellowship of the Engineering and Physical Sciences
Research Council.

\end{document}